\documentclass[a4paper,12pt]{article}
\usepackage[T1]{fontenc}
\pdfoutput=1 

\usepackage{comment}
\usepackage{tikz-cd}
\usepackage{jheppub}
\usepackage{macro}
\usepackage{amsmath,amssymb}
\usepackage{tikz}
\usepackage{pgfkeys}
\usepackage{pgfopts}
\usepackage{xcolor}
\usepackage{young}
\usepackage{youngtab}
\usepackage{ytableau}
\usepackage{titlesec}
\usetikzlibrary{decorations.pathreplacing}

\usepackage{tikz}

\titleformat{\paragraph}
{\normalfont\normalsize\bfseries}{\theparagraph}{1em}{}
\titlespacing*{\paragraph}
{0pt}{3.25ex plus 1ex minus .2ex}{1.5ex plus .2ex}

\author[a]{Dario Francia,}
\author[a]{Federico Manzoni}

\affiliation[a]{Mathematics and Physics Department, Roma Tre \& INFN Sezione di Roma Tre}

\emailAdd{ \\ dario.francia@uniroma3.it,  federico.manzoni@uniroma3.it, }

\title{\textbf Asymptotic charges of $p-$forms and their dualities in any $D$}

\abstract{We compute the surface charges associated to $p-$form gauge fields in arbitrary spacetime dimension for large values of the radial coordinate. 
In the critical dimension where radiation and Coulomb falloff coincide we find asymptotic charges involving asymptotic parameters, i.e.~parameters with a component of order zero in the radial coordinate. However, in  different dimensions we still find nontrivial  asymptotic charges now involving parameters that are not asymptotic times the radiation-order fields. 
For $p$=1 and $D>4$, our charges thus differ from those presented in the literature. We then show that under Hodge duality ``electric'' charges for $p-$forms are mapped to ``magnetic'' charges for the dual $q-$forms, with $q = D-p-2$. For charges involving fields with radiation falloffs the duality relates charges that are finite and nonvanishing. For the case of Coulomb falloffs, above or below the critical dimension, Hodge duality exchanges overleading charges in one theory with subleading ones in its dual counterpart. 
}

\begin{document} 
\maketitle
%\flushbottom
%\newpage
\section{Introduction}
Gauge theories involving  $p-$form fields emerge naturally in many physical contexts. In string theory, D$p-$branes can be introduced as sources carrying the charge of $(p+1)-$forms while $p-$form-type massive particles provide the simplest specimens of the ``exotic'' part of the string spectrum containing mixed-symmetry representations of $SO(24)$ or $SO(8)$. In supersymmetric and supergravity theories, irreducible multiplets  generically contain $p-$forms \cite{Henneaux:1986ht,green_schwarz_witten_2012, cecotti_2015}. More recently, in the context of celestial holography,  a basis of conformal primary wave functions for these fields was constructed in \cite{Donnay:2022ijr}.

The on-shell degrees of freedom of a $p-$form gauge field in dimension $D$ are carried by two equivalent irreducible  representations of $SO(D-2)$, given by antisymmetric $SO(D-2)$ tensors of rank $p$ and $D-p-2$, respectively. The corresponding covariant field theories thus provide dual realizations of the same physical degrees of freedom, while still being off-shell inequivalent in terms of the corresponding irreps of $GL(D)$ (see {\it e.g.}~\cite{hamermesh1989group}). In particular, whereas the gauge groups of dual theories are generically different, there could be in principle an interesting interplay between the corresponding asymptotic charges given that the latter are evaluated on-shell. The simplest example to this effect is provided by the infinitely-many asymptotic charges involving a massless scalar field first proposed in $D=4$ in \cite{Campiglia_20182}, whose symmetry origin was later identified in terms of the  asymptotic charges of the corresponding dual two-form \cite{Campiglia_2019, Francia_2018,Ferrero:2024eva}. See also \cite{Henneaux:2018mgn, Gonzalez:2024rho} for a Hamiltonian perspective on the same issue at both spatial and null infinity, and e.g. \cite{Strominger:2017zoo, McLoughlin:2022ljp, Ciambelli:2022vot, Donnay:2023mrd} for general reviews on asymptotic symmetries from different vantage points.

In this work we would like to investigate the correspondences between dual charges at null infinity holding for $p-$forms and $(D-p-2)-$forms for any $p$ and in any $D$, thus in particular generalising the interplay between electric and magnetic asymptotic charges highlighted for Maxwell fields in $D=4$ in \cite{Strominger_2016}.  For analogous investigations on the gravitational side see \cite{Hull:2023iny}. We evaluate the surface charges on field configurations displaying two types of falloffs,
\begin{align}
 \varphi_{r} &\sim {\cal O} (r^{-\frac{D-2}{2}}), \qquad & \mbox{radiation falloffs,} \label{rfo}\\
 \varphi_C &\sim {\cal O} (r^{-(D-p-2)}), \qquad & \mbox{Coulomb falloffs,} \label{Cfo}
 \end{align}
that are of special physical significance: the first ones bring about information related to the radiation flux  and display a universal scaling  for any $p$, while the second ones encode the asymptotic scaling of static fields generated by extended ``electric'' sources. For any value of $p$ these falloffs coincide in the ``critical dimension'' 
\begin{equation}
    D_c = 2p + 2,
\end{equation}
while the radiation order is leading for  $D>D_c$ and subleading for $D<D_c$.  Various aspects of the asymptotic symmetry structure of $p-$forms in $D=D_c$ were explored in \cite{Afshar:2018apx, Esmaeili:2020eua}.

We work in Lorenz gauge and look for parameters that keep as much as possible \eqref{rfo}  and \eqref{Cfo}, under the condition that their leading components still display an arbitrary dependence on the angular coordinates. To this end, consistently with previous results concerning $p=1, 2$ \cite{Campiglia:2016hvg, Campoleoni:2019ptc,  Ferrero:2024eva}
we admit a polyhomogeneous expansion of the residual symmetry parameters whose first subleading term is proportional to $\ln r$. Such logarithmic contributions in their turn lead to a ``mild'' violation of the field falloffs \eqref{rfo} or \eqref{Cfo}, although in pure-gauge sectors. Proceeding in this fashion we construct two classes of charges, one involving field configurations at radiation order, that we denote $Q_p^r$, and one containing fields at Coulomb orders, $Q_p^C$, displaying the falloffs
\begin{equation}
    Q_p^r \sim {\cal O} (r^0) \, , \qquad \qquad
    Q_p^C \sim {\cal O} (r^{-(D-2p-2)}) \, .
\end{equation}
We thus obtain in any $D$  asymptotically finite charges $Q_p^r$ involving the radiation order.
Let us remark that these charges involve {\it asymptotic} parameters, {\it i.e.} parameters possessing a component of order ${\cal O} (r^0)$ in their radial expansion, only in the critical dimension $D=D_c$.
Differently, they involve components of the parameters that are overleading w.r.t. ${\cal{O}} (r^0)$ in $D<D_c$ and subleading for $D>D_c$.
For instance for the $p=1$ case of Maxwell fields our asymptotic charges in any $D$ are of the form
\begin{equation}
\begin{aligned}
    &Q_1^r = \int_{S_u} d\Omega\, \epsilon^{(\frac{D-4}{2})}(x^i) H^{(\frac{D}{2})}_{ru}\, .
    \label{charge1rad}
\end{aligned}    
\end{equation}
In the expression above  $H^{(\frac{D}{2})}_{ru}$ denotes the relevant coefficient of the Maxwell field strength in Bondi coordinates   at order ${\cal O} (r^{-\tfrac{D}{2}})$, while $\epsilon^{(\frac{D-4}{2})}(x^i)$  is the coefficient of the parameter at  order  ${\cal O} (r^{-\tfrac{D-4}{2}})$ that thus describes a sector of the gauge transformations that 
diverges in $D<4$ and 
dies off, as one approaches null infinity, for $D>4$.

When taking into account the Coulomb branch for the field falloffs, instead, one finds that the charges themselves $Q_p^C$ are overleading if $D<D_c$ and subleading if $D>D_c$, while they coincide with those built out of radiation falloffs for $D=D_c$.

We then consider the dualities between charges obtained for a given $p-$form and for the corresponding form of degree $q = D-p-2$. We explicitly define the map connecting the ``electric'' charges for a given form degree to the ``magnetic'' ones of its dual counterpart via Hodge duality. Charges involving fields at the radiation order result to be of ${\cal O}(r^0)$ in both electric and magnetic incarnations, while the Coulombic charges that are divergent in a given picture get mapped to subleading charges in the dual picture, and viceversa. This latter observation indicates that for a given $p-$form gauge field  in $D>D_c$, although from the ``electric'' viewpoint the leading order for the charges is the radiation one, there will exist anyway magnetic charges that are overleading with respect to $Q^r_p$, that can be obtained as the dual charges of the subleading Coulomb charges of the dual form of degree $q = D-p-2$. Assuming that these charges  can be consistently renormalised \cite{Freidel:2019ohg, Campoleoni:2023eqp, Peraza:2023ivy}, their presence suggests that the asymptotic symmetry group of a given theory in general needs to be defined taking into account all possible physically equivalent formulations of the same theory, in particular whenever the gauge groups turn out to be different in different incarnations. 

The paper is organized as follows: in Section \ref{sec:residual} we compute the residual symmetries of $p-$forms in Lorenz gauge, that we exploit in Section \ref{sec:asymp_Q} to compute the corresponding charges for fields at radiation and at Coulomb orders. In Section \ref{sec:dualitymap} we construct our duality map for each of the two classes of charges and observe in particular the appearance of overleading magnetic charges for any $p$, when $D>D_c$. In the appendices we collect a few technical results. In particular in Appendix \ref{Appendix:Physical_logs} we provide the asymptotic analysis of the solution space and discuss the possibility to attach a physical meaning to the logarithmic terms violating the standard falloff prescription.

%%%
\section{Residual symmetries of $p-$forms in Lorenz gauge} \label{sec:residual}
%%%

%%
\subsection{Generalities}
The Lagrangian of a $p-$form gauge field $B_{\mu_1\cdots \mu_{p}}$ is 
\cite{Henneaux:1986ht}
\begin{equation} \label{LH}
    \mathcal{L}=-\frac{\sqrt{|g|}}{2 (p+1)!}H_{\mu_1\cdots\mu_{p+1}}H^{\mu_1\cdots\mu_{p+1}}
\end{equation}
where 
\begin{equation}
H_{\mu_1\cdots\mu_{p+1}}=\partial_{[\mu_1}B_{\mu_2\cdots\mu_{p+1}]}
\end{equation}
is the $(p+1)-$form field strength\footnote{Here we denote with square brackets antisymmetrization of indices with no normalization factors implied, {\it e.g.} $H_{\mu_1 \mu_2 \mu_3} = \partial_{\mu_1}B_{\mu_2\mu_3} + \partial_{\mu_3}B_{\mu_1\mu_2} + \partial_{\mu_2}B_{\mu_3\mu_1}.$}. The equations of motion are 
\begin{equation}
\partial^{\mu_{1}}H_{\mu_{1}\cdots\mu_{p+1}}= 0.
\end{equation}
The Lagrangian \eqref{LH} is invariant under
\begin{equation}\label{gaugevar}
    \delta B_{\mu_1\cdots\mu_p}=\partial_{[\mu_1}^{\phantom{(}}\epsilon^{(p-1)}_{\mu_2\cdots\mu_p]}, 
\end{equation}
where in order to identify the linearly independent gauge transformations one has to take into account the existence of a chain of $p$ gauge-for-gauge redundancies of the form
\begin{equation} \label{gaugeforgauge}
   \delta \epsilon^{(k)}_{\mu_1\cdots\mu_k}=\partial_{[\mu_1}^{\phantom{(}}\epsilon^{(k-1)}_{\mu_2\cdots\mu_k]}, \qquad k=0, \ldots p-1\, . 
\end{equation}
We introduce Bondi coordinates $(u,r, x^i)$ where $u=t-r$ and  $x^i$ is a set of $D-2$ angular variables parameterizing the $(D-2)-$dimensional unit sphere $S^{D-2}$. The Minkowski line element is
\begin{equation}
    ds^2=-du^2-2dudr+r^2\gamma_{ij}dx^idx^j \ \ \ i,j=1,\ldots,D-2;
\end{equation}
where $\gamma_{ij}$ is the metric on $S^{D-2}$ while the nonvanishing Christoffel symbols are 
\begin{equation}
\Gamma^i_{jr}=\Gamma^i_{rj}=\frac{1}{r}\delta^i_j, \ \ \ \Gamma^u_{ij}=-\Gamma^r_{ij}=r\gamma_{ij}, \ \ \ \Gamma^k_{ij}=\frac{1}{2}\gamma^{kl}(\partial_j\gamma_{li}+\partial_i\gamma_{jl}-\partial_l\gamma_{ij}).
    \label{cribon}
\end{equation}
In these coordinates tensor components carry one additional power of $r$ for each index taking values on the angular coordinates, when expressed in terms of their Cartesian counterparts. 
We focus on two types of falloffs, namely radiation and Coulomb falloffs. For a $p-$form in Minkowskian coordinates they are identified with the following scalings in $r$ of each field component \cite{Afshar:2018apx}:
\begin{align}
 B \, &\sim \, {\cal O} (r^{-\tfrac{D-2}{2}}) \qquad  &\mbox{radiation falloffs,} \label{rad} \\
  B \, &\sim \, {\cal O} (r^{-(D-p-2)})  \qquad &\mbox{Coulomb falloffs.} \label{C}
\end{align}
These falloffs coincide in the ``critical'' dimension $D_c = 2 p+2$, whereas radiation falloffs are leading for $D > D_c$ and subleading for $D < D_c$. 
We work in Lorenz gauge
\begin{equation} \label{Lorenz}
    \nabla^{\alpha} B_{\alpha \mu_2 \cdots \mu_p} \, = \, 0 \, ,
\end{equation}
so that the gauge fields and the residual parameters are to satisfy\footnote{In order to reach the conditions \eqref{waveparameter} one has to perform a nested chain of Lorenz-like gauge-for-gauge fixings exploiting \eqref{gaugeforgauge}.}
\begin{equation}
\begin{split} \label{waveparameter}
   & \Box \, B_{\mu_1\cdots\mu_p} \, = \, 0 \, , \\
   & \Box \, \epsilon_{\mu_1\cdots\mu_{p-1}} \, = \, 0 \, , \\
   & \nabla^{\alpha} \epsilon_{\alpha \mu_2 \cdots \mu_{p-1}} \, = \, 0\, .
\end{split}
\end{equation}
We would like to determine the gauge parameters that keep either \eqref{rad} or \eqref{C}, while also retaining an arbitrary dependence on the angular coordinates at the corresponding leading order. However, for $p=1, 2$ in the Lorenz gauge \eqref{Lorenz} those two requirements cannot be strictly satisfied simultaneously: in order to preserve an arbitrary dependence on the angular coordinates at leading order one has to expand the parameters polyhomogeneously in $r$, so as to include in particular a first subleading term proportional to $\ln r$. See Appendix~\ref{Appendix-logs} for a proof for $p=1, 2$ in any $D$. The latter, in its turn, will induce 
%a violation of 
overleading terms compared to some of the field falloffs \eqref{rad} and \eqref{C}. One can always interpret these violations as taking place in pure gauge sectors of the theory, with no impact on observable quantities. In this work we assume the same mechanism and interpretation to hold for any $p$. 
In Appendix~\ref{Appendix:Physical_logs}, however, we take a different perspective and analyse the effect of allowing the same logarithmic falloffs in the physical sector, as additional contributions to the radial behaviour of the field strength that do not  spoil anyway the finiteness of physical quantities, like the energy flux through null infinity, thanks to cancellations holding due to the equations of motion and to the Lorenz gauge.

\subsection{Residual symmetries}
Let us consider first the example of a spin-one gauge field. With respect to the analysis in \cite{Campoleoni:2020ejn,Campoleoni:2019ptc}, here we work in any $D$ and not just even $D$. The radiation falloffs in Bondi coordinates  are
\begin{equation} \label{radfalloff1}
B_{\mu}  \sim \begin{cases}
        O\big(r^{-\frac{D-2}{2}}\big)& \quad \text{if } \quad \mu \in \{r,u\},\\
        O\big(r^{-\frac{D-4}{2}}\big)& \quad \text{if } \quad \mu=i,
    \end{cases}    
\end{equation}
while for the parameter we assume, as anticipated, a polyhomogeneous expansion of the general form
\begin{equation} \label{expansion1form}
    \epsilon=\sum_{l \in \mathbb{Z}}\frac{\epsilon^{(l)}(u, x^i)+\tilde{\epsilon}^{(l)}(u, x^i)\ln(r)}{r^l}.
\end{equation}
Where $l\in  \mathbb{Z}$  for even $D$ and $l \in  \frac{1}{2}\mathbb{Z}$ for odd $D$. 
Inserting the expansion \eqref{expansion1form} in the gauge transformation of \eqref{radfalloff1} and asking that the log terms appear at the first subleading order we get 
\begin{equation} \label{expansion1formg}
\epsilon =\frac{\epsilon^{(\frac{D-4}{2})}(x_i)}{r^\frac{D-4}{2}}+ \sum_{l\geq \frac{D-3}{2}}\frac{\epsilon^{(l)}(u,x_i)+\tilde{\epsilon}^{(l)}(u,x_i)\ln(r)}{r^l}\, ,
\end{equation}
where the $u-$dependent subleading coefficient $\tilde{\epsilon}^{(\frac{D-2}{2})}(u,x_i)$, instrumental to preserve the angular dependence of the leading component of the parameter, gives rise to a violation of the falloff for the component $B_u$ at  ${\cal{O}} \big(r^{-\frac{(D-2)}{2}}\ln(r)\big)$. We discuss the role of those logarithmic terms in Appendix~\ref{Appendix-logs}, where we also show that $\epsilon^{(\frac{D-4}{2})}(x)$ and $\epsilon^{(\frac{D-2}{2})}(u,x)$ are arbitrary functions of their arguments. Barring the additional considerations presented in Appendix ~\ref{Appendix:Physical_logs}, in the text, we take the general attitude that the corresponding falloff violations are pure large gauge transformations\footnote{Parameters leading to more drastic violations of the field falloffs in pure gauge sectors have been considered in \cite{Romoli:2024hlc} and allow to construct hierarchies of overleading charges.}. 

If one considers Coulomb falloffs,
\begin{equation} \label{Cfalloff1}
B_\mu \sim \begin{cases}
        O\big(r^{-(D-3)}\big)& \quad \text{if } \quad \mu \in \{r,u\},\\
        O\big(r^{-(D-4)}\big)& \quad \text{if } \quad \mu=i,
    \end{cases}    
\end{equation}
assuming again an expansion as in \eqref{expansion1form}, one finds
\begin{equation} \label{expansion1form1}
\begin{aligned}
\epsilon=\frac{\epsilon^{({D-4})}(x_i)}{r^{D-4}}+\sum_{l\geq {D-3}}\frac{\epsilon^{(l)}(u,x_i)+\tilde{\epsilon}^{(l)}(u,x_i)\ln(r)}{r^l}.
\end{aligned}
\end{equation}
Let us mention that in this case equations of motion and guage fixing do not suffice to cancel out the logarithmic overleading terms in $H_{ru}$ that would appear upon assuming {\it physical} logarithmic corrections to the first of \eqref{Cfalloff1}. The corresponding contributions would still be subleading in $D>D_c$ but would be relevant for $D<D_c$, where the Coulombic branch is the leading one. Thus, for this latter case it seems that the only consistent option is to interpret them as pure gauge contributions. 

For the two-form the radiation falloffs are 
\begin{equation}
B_{\mu \nu} \sim \begin{cases}
        O\big(r^{-\frac{(D-2)}{2}}\big)&  \quad \text{if } \quad \mu, \nu \in \{r,u\}\, ,\\
        O\big(r^{-\frac{(D-4)}{2}}\big)&  \quad \text{if } \quad \mu, \nu \in \{r,i\} \ \text{or} \ \{u,i\}\, ,\\
        O\big(r^{-\frac{(D-6)}{2}}\big)& \quad \text{if } \quad   \mu, \nu \in \{i,j\}\, .
    \end{cases}
\end{equation}
We assume an expansion as in \eqref{expansion1form}, with a vector parameter, and find, after some algebra
\begin{equation} \label{parameter2rad}
\epsilon_j=\frac{\epsilon_{j}^{(\frac{D-6}{2})}(x^i)}{r^\frac{D-6}{2}}+\sum_{l\geq \frac{D-5}{2}}\frac{\epsilon_{j}^{(l)}(u, x^i)+\tilde{\epsilon}_{j}^{(l)}(u,x^i)\ln(r)}{r^l},
\end{equation}
Where we took into account the conditions for the residual gauge symmetry 
\begin{equation}
    \Box \epsilon_{\alpha}=\Box \epsilon=\nabla^{\mu}\epsilon_{\mu}=0 .
\end{equation}
with $\epsilon$ denoting the gauge-for-gauge scalar parameter. Let us observe  in particular that we  employed the latter to perform the gauge-for-gauge fixing
\begin{equation}
    \tilde{{\epsilon}}_u^{(\frac{D}{2})}= 
    \epsilon_j^{(\frac{D-8}{2})}= 
    \tilde{\epsilon}_j^{(\frac{D-6}{2})}=0.
    \label{gxg}
\end{equation}
Starting instead from Coulombic falloffs
\begin{equation}
B_{\mu \nu} \sim \begin{cases}
        O\big(r^{- (D-4)}\big)&  \quad \text{if } \quad \mu, \nu \in \{r,u\}\, ,\\
        O\big(r^{-(D-5)}\big)&  \quad  \text{if } \quad \mu, \nu \in \{r,i\} \ \text{or} \ \{u,i\}\, ,\\
        O\big(r^{- (D-6)} \big)& \quad \text{if }  \quad  \mu, \nu \in \{i,j\}\,,
    \end{cases}
\end{equation}
one gets
\begin{equation} \label{parameter2C}
\epsilon_i=\frac{\epsilon_{i}^{({D-6})}(x^i)}{r^{D-6}}+\sum_{l\geq {D-5}}\frac{\epsilon_{i}^{(l)}(u,x^i)+\tilde{\epsilon}_{i}^{(l)}(u,x^i)\ln(r)}{r^l}.
\end{equation}
Let us also mention that for both radiation and Coulomb falloffs it is possible to show that, in dimension $D>D_c$, possible overleading components in $\epsilon_j (x^i)$ can be set to zero performing gauge-for-gauge fixings. 

For a $p-$form the analysis is essentially the same with more options for the angular indices,
%
%\begin{equation}
%\begin{aligned}
%B_{ru\underbrace{ij\cdots k}_{p-2}}, \ \ B_{r\underbrace{ij\cdots k}_{p-1}}, \ \ B_{u\underbrace{\footnotesize ij\cdots k}_{p-1}}, \ \ B_{\underbrace{ij\cdots k}_{p}},
%\end{aligned}    
%\end{equation}
%
\begin{equation}
\begin{aligned}
B_{u r i_1 i_2\cdots i_{p-2}}\,, \ \ B_{r i_1 i_2\cdots i_{p-1}}, \ \ B_{u i_1 i_2\cdots i_{p-1}}\,, \ \ B_{i_1 i_2\cdots i_p}\,,
\end{aligned}    
\end{equation}
and a bit more algebraic manipulations required, and where the asymptotic expansions of the various components of the field are given in \eqref{ansatzp}.  However, if we focus on $\epsilon_{i_1\cdots i_{p-1}}$, which is  the parameter that eventually enters the asymptotic charge, in the polyhomogeneous expansion 
\begin{equation} \label{expansionformflin}
    \epsilon_{i_1\cdots i_{p-1}}=\sum_{l\in \mathbb{Z}}\frac{\epsilon_{i_1\cdots i_{p-1}}^{(l)}(u,x^i)+\tilde{\epsilon}_{i_1\cdots i_{p-1}}^{(l)}(u,x^i)\ln(r)}{r^l},
\end{equation}
the leading order is in general a linear function $f(p,D)$ of $p$ and $D$. As such, we can determine its general form for both radiation and Coulombic falloffs making use of the results found for $p=1, 2$, finding
\begin{equation} \label{f_pD}
\begin{aligned}
    &f_r (p,D)=\frac{D-(2p+2)}{2}, \\
    &f_C(p,D) =D-(2p+2).
\end{aligned}    
\end{equation}
\section{Asymptotic charges} \label{sec:asymp_Q}
%%%
Following Noether's second theorem \cite{Avery_2016} 
the charges take the form
\begin{equation} \label{Q}
Q =\int_{\Sigma}j^{\mu}d\Sigma_{\mu}=\int_{\partial \Sigma} k^{\mu \nu} d\sigma_{\mu \nu},
\end{equation}
where in the case of interest for us we integrate over the sphere $S^{D-2}$ at null infinity, {\it i.e.} keeping $u$ and $r$ fixed, and then taking the limit $r \rightarrow +\infty$. From the variation of the Lagrangian \eqref{LH} one finds, up to factors
\begin{equation}
    k^{\mu \nu} = \epsilon_{\rho_1\cdots \rho_{p-1}}H^{\mu \nu \rho_1\cdots \rho_{p-1}}.
\end{equation}
In Bondi coordinates the integration measure in \eqref{Q} brings about a factor of $r^{D-2}$. Thus, in order for the charges to have a finite and nonvanishing limit for $r\to \infty$, one needs 
\begin{equation} \label{order_k}
    k^{\mu \nu} \sim {\cal O} ({r^{-({D-2)}}}).
\end{equation}
For {\it asymptotic} parameters, i.e. parameters whose radial expansion is of the schematic form 
\begin{equation}
\epsilon (u, r, x^i) = \epsilon^{(0)} (x^i) + \mbox{subleading terms},
\end{equation}
at least for the relevant components of $\epsilon$ entering \eqref{Q}, the corresponding components of the field strength are to satisfy $H \sim {\cal O} ({r^{-({D-2)}}})$. However, finite asymptotic charges can also involve non-asymptotic parameters, as long as \eqref{order_k} holds. 

In the following we will compute the formal charges that involve the leading parameters found in the previous section for radiation and Coulomb falloffs, respectively. For the case of radiation falloffs we always find finite asymptotic charges, involving however full-fledged asymptotic parameters only in the critical dimension. For the case of Coulomb falloffs in $D\neq D_c$ the corresponding charges are either overleading, if $D<D_c$, or subleading, if $D>D_c$. As we shall see in Section~\ref{sec:dualitymap} all these charges turn out to be relevant in the discussion of dualities between $p-$forms and $q=(D-p-2)-$forms. In our writing of the charges we will consistently omit overall factors.

\subsection{$1-$forms}
The asymptotic charge in the case of $1-$forms is
\begin{equation}     \label{carica1forma}
    Q_1 = \lim_{r \to +\infty}\int \epsilon H_{ru}r^{D-2}d\Omega.
\end{equation}
For the case of radiation falloff we see from \eqref{expansion1formg} that the leading order term of the parameter scales like ${\cal O} (r^{\frac{4-D}{2}})$, so that the leading contribution to the charge is given by
\begin{equation}   \label{charge1rad}
    Q^r_1 = \int \epsilon^{(\frac{D-4}{2})}(x^i) H^{(\frac{D}{2})}_{ru} d\Omega,
\end{equation}
where 
\begin{equation}  \label{Hterms_1}
   H^{(\frac{D}{2})}_{ru}=-\left(\frac{D-2}{2}\right)B_{u}^{(\frac{D-2}{2})}-\partial_uB_{r}^{(\frac{D}{2})}\, .
\end{equation}
The charge \eqref{charge1rad} is indeed asymptotic, i.e. of order  ${\cal O} (r^0)$.
To make \eqref{Hterms_1} more explicit, following the analysis in \cite{Campoleoni:2019ptc}, we can set $B_u^{(\frac{D-2}{2})}$ to zero by exploiting the arbitrary $u-$dependence of $\epsilon^{(\frac{D-2}{2})}$ (see the discussion below \eqref{eq:explicitsoleps}). Note that this amounts to a small gauge transformation because this component of the gauge parameter does not appear in the asymptotic charge \eqref{charge1rad}. Let us also note that the Lorenz gauge condition combined with the $r-$equation of motion (see Eqs.~\eqref{divfreepform} and \eqref{orbyorpr} for $p=1$, without fields with tilde superscript)  impose
\begin{equation}
\begin{aligned}
\partial_u B_r^{(l)} &= (l-D+1)  (B_u^{(l-1)}
    -B_r^{(l-1)})
    + \nabla^iB_i^{(l-2)}\,,\\
    (D-2l-2) \partial_u B_r^{(l)} &= [
\Delta + l(l-D+1)  
    ]
    B_r^{(l-1)}
    +(D-2)B_u^{(l-1)} - 2 \nabla^iB_i^{(l-2)}\,.
\end{aligned}
\end{equation}
In view of the above conditions, for $l=\frac{D}{2}$, combining the two equations imposes  $B_r^{(\frac{D-2}{2})}=0$, while the first one  yields
\begin{equation}
\partial_u B_r^{(\frac{D}{2})} = 
    \nabla^iB_i^{(\frac{D-4}{2})}\,.
\end{equation}
Substituting into \eqref{charge1rad}, we finally obtain
\begin{equation}\label{eq:chargespin1radexplicit}
     Q^r_1 = -\int \epsilon^{(\frac{D-4}{2})}(x^i) \nabla^iB_i^{(\frac{D-4}{2})}(u,x^i) d\Omega,
\end{equation}
This result matches the one in footnote~4 of \cite{Campoleoni:2019ptc}.
This charge is intrisically related to the presence of radiation encoded in the field component $B_i^{(\frac{D-4}{2})}$, which indeed is the relevant one to calculate the energy flux through null infinity (see Eq.~(3.11) of \cite{Campoleoni:2019ptc}).

For Coulomb fall-offs, given
the radial behaviour of the parameter in \eqref{expansion1form1}, one finds 
\begin{equation}     \label{carhe1coul}
 Q^C_1 = \lim_{r \to +\infty}\int \frac{\epsilon^{({D-4})}(x^i)}{r^{D-4}}\bigg[H_{ru}^{(D-2)}+\tilde{H}_{ru}^{(D-2)}\ln(r)\bigg]d\Omega.
\end{equation}
In this expression we also considered the possibility that the logarithmic terms in the polyhomogeneous expansion \eqref{expansion1form1} be associated to corresponding falloffs in the physical sector of the theory, so that they are present in the radial expansion of the field strength. Differently from the case of the energy flux for radiation falloffs, discussed in Appendix~\ref{Appendix:Physical_logs}, we see that such terms would indeed affect the charge and produce a leading Coulombic behaviour $\sim 
{\cal O} (r^{-(D-4)} \ln r)$.
From the equations of motion and the gauge condition one finds
\begin{equation} \label{FS1}
\begin{aligned}
   &H_{ru}^{(D-2)}=-(D-3)B_{u}^{(D-3)}-\partial_uB_{r}^{(D-2)}+\tilde{B}_{u}^{(D-3)},\\
   &\tilde{H}_{ru}^{(D-2)}=-(D-3)\tilde{B}_{u}^{(D-3)}-\partial_u\tilde{B}_{r}^{(D-2)}=(D-4)\tilde{B}_{u}^{(D-3)}.
\end{aligned}   
\end{equation}
so that in particular, by the second of \eqref{FS1},one sees that the charge \eqref{carhe1coul} is anyway finite in the critical dimension $D=4$. 
Differently, irrespective of the log terms, \eqref{carhe1coul} it vanishes in $D>4$ and diverges in $D<4$. 

In what follows, however, we will still assume that the log terms only affect pure gauge configurations, and as such do not alter the form of the charges, even for the Coulomb branch of the field falloffs. Additional considerations on their possible physical role will be proposed in Appendix \ref{Appendix:Physical_logs}.

\subsection{$2-$forms}
In this case we need to evaluate
\begin{equation}  \label{carica2forma}
Q_2 = \lim_{r \rightarrow +\infty}\int\gamma^{ij}\epsilon_iH_{ruj}r^{D-4}d\Omega;
\end{equation}
where
\begin{equation}     \label{carica2formaes}
\begin{split}
H_{ruj} & =  \nabla_rB_{uj}-\nabla_uB_{rj}-\nabla_{j}B_{ur} \\
   & = \partial_rB_{uj}-\frac{1}{r}B_{uj}-\partial_uB_{rj}-\partial_{j}B_{ur}+\frac{1}{r}B_{uj}
\end{split}
\end{equation}
For the case of radiation falloffs,
the gauge parameter scales as in \eqref{parameter2rad} and consequently one can find an asymptotic charge given by
\begin{equation}     \label{charge2rad}
    Q^r_2 = \int \gamma^{ij} \epsilon_{i}^{(\frac{D-6}{2})}(x^i)H_{ruj}^{(\frac{D-2}{2})}d\Omega,
\end{equation}
where 
\begin{equation}
    H_{ruj}^{(\frac{D-2}{2})}=-\left(\frac{D-4}{2}\right)B_{uj}^{(\frac{D-4}{2})}-\partial_{j}B_{ur}^{(\frac{D-2}{2})}-\partial_uB_{rj}^{(\frac{D-2}{2})}.
\end{equation}
The same kind of analysis performed for the one-form case can be repeated in this setting and shows that \eqref{charge2rad} depends on the free radiation data. However, we postpone the discussion to the general case of $p-$forms in the next paragraph. 

Differently, for Coulomb falloffs
the parameter in \eqref{parameter2C} to leading order scales like ${\cal{O}} (r^{6-D})$ 
and thus the leading contribution to the corresponding charge is given by
\begin{equation}     \label{carhe2coul}
\begin{aligned}
    &Q^C_2 = \lim_{r \to +\infty}\int \gamma^{ij} \frac{\epsilon_{i}^{({D-6})}(x^i)}{r^{D-6}} H_{ruj}^{(D-4)} d\Omega,
\end{aligned}    
\end{equation}
where
\begin{equation}
\begin{aligned}
&H_{ruj}^{(D-4)}=-(D-5)B_{uj}^{(D-5)}-\partial_{j}B_{ur}^{(D-4)}-\partial_uB_{rj}^{(D-4)};\\
\end{aligned}   
\end{equation}
One can notice that in $D=4$ the radiation charge \eqref{charge2rad} involves an overleading parameter \cite{Ferrero:2024eva, Romoli:2024hlc}. The Coulombic charge \eqref{carhe2coul} in its turn vanishes in $D>6$ and diverges in $D<6$. In $D=6$ \eqref{charge2rad} 
coincides with \eqref{charge2rad}. 

\subsection{$p-$forms}\label{chargecomputation}
The general case reproduces what seen above up to a somewhat  heavier algebra. Barring an overall factor the surface charge is given by
\begin{equation}
\begin{aligned} \label{caricapforma}
    Q_p=&\lim_{r \rightarrow +\infty}\int r^{D-2p} \gamma^{i_1j_1}\cdots \gamma^{i_{p-1}j_{p-1}}\epsilon_{i_1\cdots i_{p-1}}H_{ru j_1\cdots j_{p-1}} d\Omega \\
    =&\lim_{r \rightarrow +\infty}\int r^{D-2p}\gamma^{i_1j_1}\cdots \gamma^{i_{p-1}j_{p-1}}\epsilon_{i_1\cdots i_{p-1}}\\
    & \bigg[\partial_{r}B_{uj_1\cdots j_{p-1}}-\partial_{u}B_{rj_1\cdots j_{p-1}}-\sum_{j=1}^{p-1}\partial_{j_j}B_{u\cdots j_{j-1}rj_{j+1}\cdots j_{p-1}}\bigg]d\Omega.
%\label{caricapformaes}
\end{aligned}
\end{equation}
For radiation falloffs, according to  \eqref{expansionformflin} and \eqref{f_pD} the radial dependence of the gauge parameter we are interested in is described by the expansion
\begin{equation}
\begin{aligned}
\epsilon_{i_1\cdots i_{p-1}}&=\frac{\epsilon_{i_1\cdots i_{p-1}}^{(\frac{D-(2p+2)}{2})}(x^i) }{r^\frac{D-(2p+2)}{2}}+\sum_{l\geq \frac{D-(2p+1)}{2}}\frac{\epsilon_{i_1\cdots i_{p-1}}^{(l)}(u, x^i)+\tilde{\epsilon}_{i_1\cdots i_{p-1}}^{(l)}(u, x^i)\ln(r)}{r^l},
\end{aligned}
\end{equation}
which determines an asymptotic charge of the form
\begin{equation} \label{chargeprad}
\begin{aligned}
    Q^r_p = \int \gamma^{i_1j_1}\cdots \gamma^{i_{p-1}j_{p-1}} \epsilon_{i_1\cdots i_{p-1}}^{(\frac{D-(2p+2)}{2})}(x^i) H_{ruj_1\cdots j_{p-1}}^{(\frac{D-2p+2}{2})} d\Omega,
\end{aligned}
\end{equation}
where
\begin{equation} \label{Fieldstrength_p}
 H_{ruj_1\cdots j_{p-1}}^{(\frac{D-2p+2}{2})}=-\bigg(\frac{D-2p}{2}\bigg)B_{uj_1\cdots j_{p-1}}^{(\frac{D-2p}{2})}-\sum_{k=1}^{p-1}(-1)^k\partial_{j_k}B^{(\frac{D-2p+2}{2})}_{ru\cdots j_{k-1}j_{k+1}\cdots j_{p-1}}-\partial_uB_{rj_1\cdots j_{p-1}}^{(\frac{D-2p+2}{2})} .
\end{equation}
Let us stress that 
\begin{equation}
    Q^r_p \sim {\cal{O}} (r^0), 
\end{equation}
i.e. the charges \eqref{chargeprad} are asymptotic for every $p$ in every $D$; in dimension $D=2p$, similarly to the case of the 1-form and the 2-form the first term is not present.
In \eqref{Fieldstrength_p} we can gauge fix $B_{rui_i...i_{p-2}}^{(\frac{D-2p+2}{2})}$ to zero by exploiting  $\epsilon_{ri_1...i_p-2}^{(\frac{D-2p+2}{2})}$.  Note that this amounts to a small gauge transformation because these components of the gauge parameter does not appear in the asymptotic charge \eqref{chargeprad}. Moreover, the Lorenz gauge condition combined with the appropriate component of the equation of motion (see Eqs.~\eqref{divfreepform} and \eqref{orbyorpr}, and the nearby definitions for the coefficients) impose
\begin{equation}
\begin{aligned}
\partial_uB_{ri_1...i_{p-1}}^{(l)}&=G_{p,D}(B_{u{i_1...i_{p-1}}}^{(l-1)}-B_{ri_1...i_{p-1}}^{(l-1)})+\nabla^iB_{ii_1...i_{p-1}}^{(l-2)}\,,\\
-A_1\partial_uB^{(l)}_{ri_1\cdots i_{p-1}}&=[k_r^{(p)}(l)+\Delta]B^{(l-1)}_{ri_1\cdots i_{p-1}}+A_2 B^{(l-1)}_{ui_1\cdots i_{p-1}}+\sum B^{(l-1)}-2\nabla^iB^{(l-2)}_{ii_1\cdots i_{p-1}}\,.
\end{aligned}
\end{equation}
In view of the above conditions, for $l=\frac{D-2p+2}{2}$, combining the two equations imposes that
\begin{equation}
\partial_u B_{ri_1...i_{p-1}}^{(\frac{D-2p+2}{2})} = 
    \nabla^iB_{ii_1...i_{p-1}}^{(\frac{D-2p}{2})}\,,
\end{equation}
up to possible non-dynamical terms given by eigenfunctions of the Laplacian on the celestial sphere. Substituting into \eqref{chargeprad}, we finally obtain
\begin{equation} \label{charges_p_rad}
     Q^r_p = -\int \gamma^{i_1j_1}\cdots \gamma^{i_{p-1}j_{p-1}} \epsilon_{i_1\cdots i_{p-1}}^{(\frac{D-(2p+2)}{2})}(x^i) \nabla^iB_{ij_1...j_{p-1}}^{(\frac{D-2p}{2})} d\Omega,
\end{equation}
thus showing that the charge involves the radiation data of the gauge field.

Looking at Coulomb falloffs we focus instead on gauge parameters displaying the expansion 
\begin{equation}
\begin{aligned}
\epsilon_{i_1\cdots i_{p-1}}&=\frac{\epsilon_{i_1\cdots i_{p-1}}^{({D-(2p+2)})}(x^i) }{r^{D-(2p+2)}}+\sum_{l\geq {D-(2p+1)}}\frac{\epsilon_{i_1\cdots i_{p-1}}^{(l)}(u, x^i)+\tilde{\epsilon}_{i_1\cdots i_{p-1}}^{(l)}(u, x^i)\ln(r)}{r^l}.
\end{aligned}
\end{equation}
When inserted in \eqref{caricapforma} one finds leading and subleading orders given by 
\begin{equation}     \label{chargepcoul}
\begin{aligned}
    &Q^C_p = \lim_{r \to + \infty} \int \gamma^{i_1j_1}\cdots \gamma^{i_{p-1}j_{p-1}} \frac{\epsilon_{i_1\cdots i_{p-1}}^{({D-(2p+2)})}(x^i)}{r^{D-2p -2}} H_{ruj_1\cdots j_{p-1}}^{(D-2p)} d\Omega, 
\end{aligned}
\end{equation}
where
\begin{equation}
\begin{aligned}
H_{ruj_1\cdots j_{p-1}}^{(D-2p)}= &-\bigg(D-2p-1\bigg)B_{uj_1\cdots j_{p-1}}^{(D-2p-1)}-\sum_{j=1}^{p-1}\partial_{j_j}B^{(D-2p)}_{u\cdots j_{j-1}rj_{j+1}\cdots j_{p-1}} - \partial_uB_{rj_1\cdots j_{p-1}}^{(D-2p)}.
\end{aligned}   
\end{equation}
In general
\begin{equation}
    Q^C_p \sim {\cal{O}} (r^{2p+2-D}), 
\end{equation}
so that these charges vanish in $D>2p+2$ while they display a power-law divergent behaviour in $D<2p+2$.

%%% 
\section{The duality map}\label{sec:dualitymap}
%%%
Gauge fields described by $p-$forms and $(D-p-2)-$forms provide
independent irreps of $GL(D)$, whenever they both exist, associated to gauge symmetries that are also independent. However, in force of their on-shell duality, one expects that any charge computed in one theory should have a counterpart in the dual partner theory. The true issue at stake is to explore whether Hodge-duality connecting asymptotic charges allows one to highlight sectors of the asymptotic symmetry group of a theory that are only visible, or  more easily identified, from the dual perspective. The prototypical instance of this phenomenon is provided by the dual pair (scalar field)-(two form) in $D=4$, where one can indeed construct asymptotic charges for the scalar theory but it is only through Hodge duality that the connection to asymptotic symmetries can be made apparent \cite{Campiglia_2019, Francia_2018, Henneaux:2018mgn, Ferrero:2024eva}. For the case of electromagnetism in $D=4$ the same issue was investigated in \cite{Strominger_2016}, while $p-$forms in the critical dimension in where considered in \cite{Afshar:2018apx, Esmaeili:2020eua} and in $AdS_D$ in \cite{Esmaeili:2021szb}.

Our main goal in this section is to derive a duality map between $p-$form charges and $(D-p-2)-$form charges making use of the duality map connecting the corresponding field strengths in different descriptions. We consider first the asymptotic charges \eqref{chargeprad}, displaying a well-defined and non-vanishing asymptotic limit in any $D$, to then extend our considerations to the charges \eqref{chargepcoul}, for which the duality map will relate vanishing (divergent) charges in one description to divergent (vanishing) charges in its dual counterpart. The map that we implement in this work relates electric charges in one setup to magnetic charges in its dual counterpart. An alternative procedure, meant to define a map relating electric charges to electric charges will be proposed in \cite{Manzoni}.

\subsection{Duality map at radiation order}\label{dualmap1}
Let us consider first the case of the dual pair ($1-$form)-($2-$form) in $D=5$.
The charges involving the radiation order are
\begin{equation} \label{Q12D5}
\begin{aligned}
    Q^r_1 &= \int \epsilon^{(\frac{1}{2})}(x^i)F_{ru}^{(\frac{5}{2})}d\Omega, \\
    Q^r_2 &= \int \gamma^{ij}\epsilon_{i}^{(-\frac{1}{2})}(x^i)H_{ruj}^{(\frac{3}{2})}d\Omega,
\end{aligned}
\end{equation}
where the integral is over the $3-$sphere at null infinity. We will refer to these charges as the ``electric'' charges of the two theories. We aim to exploit the Hodge-duality between the field strengths 
(we adopt the mostly-plus signature for the metric)
\begin{equation}
\begin{split}
    F &= -\star H,  \\
    H &= \star F\, ,
\end{split}
\end{equation}
that in components reads
\begin{equation} \label{dual12D5}
 \begin{split}
    F_{\alpha \beta} & =-\frac{r^3}{6}g^{\rho \tilde{\rho}}g^{\nu \tilde{\nu}}g^{\mu \tilde{\mu}}H_{\mu \nu \rho}\varepsilon_{\tilde{\mu}\tilde{\nu}\tilde{\rho}\alpha \beta}\sqrt{\gamma}, \\
      H_{\mu \nu \rho}& =\frac{r^3}{2}g^{\alpha \tilde{\alpha}}g^{\beta \tilde{\beta}}F_{\alpha \beta}\varepsilon_{\tilde{\alpha} \tilde{\beta}\mu \nu \rho}\sqrt{\gamma},
 \end{split}
\end{equation}
to show that the one-form charge in \eqref{Q12D5} can be given a meaning as a charge in the two-form theory, and viceversa. Although in this section we will be basically concerned with the duality at the level of field strengths, let us observe that from \eqref{dual12D5} one can also link the one-form leading order field components $A_{i}^{(\frac{1}{2})}$ and their dual two-form leading order field components $B_{ij}^{(-\frac{1}{2})}$, in the spirit of \cite{Strominger_2016}. In fact, from the first of \eqref{dual12D5} with $(\alpha ,\beta)=(u,i)$ we get
\begin{equation}
    F_{ui}^{(\frac{1}{2})}=-{\sqrt{\gamma}}\varepsilon_{urijk}\gamma^{ks}\gamma^{jl}H_{uls}^{(-\frac{1}{2})};
\end{equation}
that allow one to relate the asymptotic electric and magnetic vector potentials, up to an angle dependent integration constant which can be fixed in such a way that $A_{i}^{(\frac{1}{2})}$ and 
$B_{ij}^{(-\frac{1}{2})}$ be proportional.

A natural dual map between the charges can be obtained exploiting \eqref{dual12D5} in the integrands of \eqref{Q12D5}. In particular, from the first of \eqref{dual12D5} one finds
\begin{equation}\label{eq:myH}
F_{ls}^{(\tfrac{1}{2})}=-\sqrt{\gamma}\gamma^{ij}\varepsilon_{urils}H_{ruj}^{(\frac{3}{2})}.
\end{equation}
Upon identifying $\varepsilon_{urils}=\varepsilon_{ils}$ and contracting both sides of \eqref{eq:myH} with $\varepsilon_{lsk}$, one then finds
\begin{equation} \label{hodge12}
\varepsilon^{lsk}F_{ls}^{(\tfrac{1}{2})}=-2\gamma^{kj}H_{ruj}^{(\frac{3}{2})}\sqrt{\gamma}, 
\end{equation}
where we took into account that $\varepsilon^{lsk}\varepsilon_{lsi}=2\delta^k_i$. One can verify that in the space of asymptotic solution the components of the one-form field strength in \eqref{hodge12} do not vanish. Therefore, substituting \eqref{hodge12} in the second of \eqref{Q12D5} we map the electric charge for the two-form in a magnetic charge for the one-form corresponding to radiation falloffs\footnote{Note that, introducing the volume form on the sphere $\omega_{ijk} = \sqrt{\gamma} \varepsilon_{ijk}$ and $\omega^{ijk}=\gamma^{ii'}\gamma^{jj'}\gamma^{kk'}\omega_{i'j'k'}=\frac{1}{\sqrt{\gamma}}\,\varepsilon^{ijk}$, Eq.~\eqref{Map2to1radiation} can be written in the manifestly covariant form
\begin{equation} \nonumber
Q^r_2 \to \tilde{Q}^r_1 =\int_{S_u^{3}}\ \omega^{ijk}\epsilon_i^{(-\frac{1}{2})}(x^i)\,F_{jk}^{(\frac{1}{2})}d\Omega.
\end{equation}
}: 
\begin{equation} \label{Map2to1radiation}
\begin{aligned}
Q^r_2 \to \tilde{Q}^r_1 =-\int_{S_u^{3}}\frac{\varepsilon^{ijk}\epsilon_i^{(-\frac{1}{2})}(x^i)}{\sqrt{\gamma}}F_{jk}^{(\frac{1}{2})}d\Omega,
\end{aligned}    
\end{equation}
where, here and in the following, the supersctript tilde is meant to identify magnetic charges  and where we interpret the vector parameter in \eqref{Map2to1radiation} as the gauge parameter of the two-form, magnetic photon\footnote{One could also formally relate the vector parameter in \eqref{Map2to1radiation} to the scalar one of the ``electric'' photon by choosing for instance $\epsilon_i^{(-\frac{1}{2})}(x^i)=\epsilon^{(-\frac{1}{2})}(x^i)(1,1,1)$. However, this choice hides the nature of the magnetic spin-one gauge symmetry in $D=5$. In addition, in order to get the charge \eqref{Map2to1radiation} only using the electric part of the gauge symmetry \eqref{expansion1formg}, one should then extend that expansion to also include an overleading term.}.
In a similar fashion, one can  derive a magnetic charge for the two-form with radiation falloffs,  $\tilde{Q}^r_2$, upon dualising  the electric one-form charge with the same falloffs $Q^r_1$:
\begin{equation}
     \tilde{Q}^r_2 = \int_{S_u^{3}}\frac{\varepsilon^{ijk}\epsilon^{(\frac{1}{2})}(x^i)}{\sqrt{\gamma}}H_{ijk}^{(-\frac{1}{2})}d\Omega,
\end{equation}
involving this time a  subleading parameter.

In the general case we consider the electric charges
involving radiation falloffs:
\begin{equation} \label{Qpq2D}
\begin{aligned}
    &Q^r_p = \int \gamma^{i_1j_1}\cdots \gamma^{i_{p-1}j_{p-1}} \epsilon_{i_1\cdots i_{p-1}}^{(\frac{D-(2p+2)}{2})}(x^i) H_{ruj_1\cdots j_{p-1}}^{(\frac{D-2p+2}{2})} )d\Omega; \\
    &Q^r_q = \int \gamma^{i_1j_1}\cdots \gamma^{i_{q-1}j_{q-1}} \epsilon_{i_1\cdots i_{q-1}}^{(\frac{D-(2q+2)}{2})}(x^i) H_{ruj_1\cdots j_{q-1}}^{(\frac{D-2q+2}{2})}d\Omega;
\end{aligned}
\end{equation}
where  $q=D-2-p$. As before, the duality map can be derived starting from the physical information of the duality between the field strengths
\begin{equation}
\begin{split}
    H_q &= \star H_p,  \\
    H_p &= (-)^{1+(p+1)(q+1)}\star H_q\, .
\end{split}
\end{equation}
In particular, for the components involving  only angular indexes they read
\begin{equation} \label{dualpqD}
\begin{aligned}
    &\frac{H_{k_1\cdots k_{p+1}}}{r^{D-2-2(q-1)}} =(-1)^{(p+1)(q+1)} \gamma^{i_1j_1}\cdots \gamma^{i_{q-1}j_{q-1}}\varepsilon_{uri_1\cdots i_{q-1}k_1\cdots k_{p+1}}H_{ruj_1\cdots j_{q-1}}\sqrt{\gamma};\\
    &\frac{H_{k_1\cdots k_{q+1}}}{r^{D-2-2(p-1)}} =\gamma^{i_1j_1}\cdots \gamma^{i_{p-1}j_{p-1}}\varepsilon_{uri_1\cdots i_{p-1}k_1\cdots k_{q+1}}H_{ruj_1\cdots j_{p-1}}\sqrt{\gamma}.
\end{aligned}
\end{equation}
Again, looking at the Hodge dual equation with one index $u$ and the remaining ones angular indices gives us
\begin{equation}
    H_{ui_1\cdots i_p}^{(\frac{D-2p-2}{2})}=(-1)^{1+(p+1)(q+1)}\varepsilon_{uri_1\cdots i_pk_1\cdots k_q}\gamma^{j_1k_1}\cdots \gamma^{j_qk_q}H_{uj_1\cdots j_q}^{(\frac{D-2q-2}{2})}\sqrt{\gamma},
\end{equation}
implies a relation between the asymptotic electric and magnetic vector potentials, up to a angle dependent integration constant which can be fixed in such a way $B_{i_1\cdots i_p}^{(\frac{D-2p-2}{2})}$ and 
$B_{i_1\cdots i_q}^{(\frac{D-2q-2}{2})}$ be proportional. 

We would like to employ \eqref{dualpqD} to trade the integrands in \eqref{Qpq2D} for angular components of the corresponding dual field strengths. One can see that the first of \eqref{dualpqD} relates the order $D-2p -2$ on the l.h.s. with $D-2q+2$ on the r.h.s. . Making use of this relation we can map the electric charge for the $q-$form to a magnetic charge for the $p-$form:
\begin{equation} \label{hodgepq}
\begin{aligned}
Q^r_q \to \tilde{Q}^r_p =  \int\frac{\varepsilon^{i_1\cdots i_{q-1}j_1\cdots j_{p+1}}{\epsilon_{i_1\cdots i_{D-p-3}}^{(-\frac{D-(2p+2)}{2})}(x^i)}}{\sqrt{\gamma}}H_{j_1\cdots j_{p+1}}^{(\frac{D-(2p+2)}{2})} d\Omega,
\end{aligned}    
\end{equation}
where
\begin{equation}
    \epsilon_{i_1\cdots i_{q-1}}^{(\frac{D-(2q+2)}{2})} = \epsilon_{i_1\cdots i_{D-p-3}}^{-\frac{D-(2p+2)}{2}}
\end{equation}
plays the role of magnetic gauge parameter for the $p-$form. To obtain the magnetic charge $\tilde{Q}^r_q$, dual to $Q^r_p$, it is enough to exchange $p \leftrightarrow q$ in \eqref{hodgepq}. Reinstalling overall signs we have
\begin{equation}   \label{eqelecmagn}
\begin{split}
Q^r_p & =\tilde{Q}^r_q, \\
Q^r_q & =(-1)^{1+(p+1)(q+1)} \tilde{Q}^r_p.
\end{split}
\end{equation}
Following \cite{Strominger_2016}, one can formally consider complexified charges 
\begin{equation}
\begin{aligned}
\mathcal{Q}^r_p &:=Q^r_p+i\tilde{Q}^r_p,\\
\mathcal{Q}^r_q &:=Q^r_q+i\tilde{Q}^r_q,
\end{aligned}    
\end{equation}
which, as in the case of electromagnetism in $D=4$, generate a complexified $\mathfrak{u}_{\mathbf{C}}(1)$ symmetry algebra. Under duality these charges transform according to a M\"obius transformation parameterized by the matrix $A \in \mathrm{PGL}(2,\mathbb{C})$ 
\begin{equation}
    A=\begin{bmatrix}
0 & 1 \\
(-1)^{1+(p+1)(q+1)} & 0  \\
\end{bmatrix}.
\end{equation}
According to the value of $(-1)^{1+(p+1)(q+1)}s$, this transformation is a either a reflection with respect the bisector of the first quadrant if $(-1)^{1+(p+1)(q+1)}=+1$ or a rotation of $\theta=\pm\frac{\pi}{2}$, (clockwise on $\mathcal{Q}^r_q$ and counterclockwise on $\mathcal{Q}^r_p$) if $(-1)^{1+ (p+1)(q+1)} =-1$. 
\subsection{Duality map at Coulomb order}
For the charges \eqref{chargepcoul} involving field components with Coulomb fall-offs one can formally reproduce the same steps, with the proviso that one carefully takes into account the different orders entering the charges.

The Coulombic magnetic charge in the $p-$form theory dual to $Q^C_q$ is then given by
\begin{equation}
    \tilde{Q}^C_p = \lim_{r \to + \infty}\int \frac{\varepsilon^{i_1\cdots i_{q-1}j_1\cdots j_{p+1}}}{\sqrt{\gamma}}
    \frac{\epsilon_{i_1\cdots i_{q-1}}^{({D-(2q+2))}}}{r^{-(D-2p-2)}}(x^i)H_{j_1\cdots j_{p+1}}^{({D-(2p+2)})}d\Omega,
    \label{magneticcharcoul}
\end{equation}
This charge is divergent in $D>2p+2$ and vanishing in $D<2p+2$, which is the opposite behavior with respect to the charges \eqref{chargepcoul}: when the Coulombic electric charge \eqref{chargepcoul} vanishes the corresponding magnetic charge \eqref{magneticcharcoul} diverges, and viceversa. Therefore a charge that in a given theory is subleading with respect to the charge with radiation falloffs is mapped to an overleading charge in the dual description. 
One simple instance is again the dual pair  two-form and scalar feld in $D=4$. In that case the overleading $\mathcal{O}(r^2)$ 2-form charge is mapped to a subleading scalar charge of order $\mathcal{O}(r^{-2})$. Scalar charges acting at  $\mathcal{O}(r^{-2})$ appear, for example, in \cite{Hamada_2017} in connection with the pion memory effect.

%%%
\section{Outlook}
%%%
Our asymptotic charges \eqref{charges_p_rad} at radiation order,  based on non-asymptotic parameters, seemingly provide an alternative to the construction of charges based on  overleading parameters, {\it i.e.} parameters involving an ${{\cal{O}} (r^0)}$ component in any $D$, typically introduced as pure gauge cotributions in order to account for the existence of asymptotic symmetries in $D>4$ \cite{Kapec:2014zla, Kapec:2015vwa, He:2019jjk, Henneaux:2019yqq, Campoleoni:2020ejn}. In this respect it would be interesting to understand to what kind of observables, if any, should those charges be related. Their actual meaning indeed is not fully clear at a first glance, since on the one hand they would be expected to generate large gauge transformations on $\cal{I}$, but on the other hand they would seemingly do so exploiting a gauge parameter that does not get to null infinity! 

An obvious generalization of our results would involve a similar exploration for gauge fields of spin $s\geq 2$.  The most direct goal would be that of constructing a new set of putative observable asymptotic quantitities. In addition, one would like to explore the possibility that hidden sectors of the asymptotic symmetry group of gravity in $D>4$, or more generally of higher-spin fields, may be visible, or at least better apparent, if explored form the perspective of the various dual formulations admissible \cite{deMedeiros:2002qpr, Boulanger:2003vs}.

%%%
\section*{Acknowledgments}
We are especially grateful to C. Heissenberg, for collaboration on the topics of this work at various stages and for valuable comments on the draft. D.F. also gratefully acknowledges useful discussions with A. Campoleoni.
%%%

\appendix
\section{Lorenz  gauge for $p=1, 2$} \label{Appendix:eqs}
The Laplace-Beltrami operator in Bondi coordinates reads
\begin{equation}
\Box=g^{\mu \nu}\nabla_{\mu}\nabla_{\nu}=-2\nabla_u\nabla_r+\nabla_r\nabla_r+\frac{1}{r^2}\gamma^{ij}\nabla_i\nabla_j\,.
\end{equation}
Let us begin by discussing the scalar gauge-for-gauge parameter. The homogeneous wave equation reads
\begin{equation}
-2\nabla_u\nabla_r\epsilon+\nabla_r\nabla_r\epsilon+\frac{1}{r^2}\gamma^{ij}\nabla_i\nabla_j\epsilon=0\,.
\end{equation}
Making explicit the covariant derivatives we get
\begin{equation}
\bigg[\partial_r^2-2\partial_u \partial_r-\frac{(D-2)}{r}(\partial_u-\partial_r)+\frac{\Delta}{r^2}\bigg]\epsilon=0,
\end{equation}
where $\Delta:=\Delta_{\mathbb{S}^{D-2}}$ is the Laplace-Beltrami operator on the $(D-2)-$sphere. Inserting a polyhomogeneous expansion gives us
\begin{equation}
\begin{aligned}
&[(l-1)(l-D+2)+\Delta]\tilde{\epsilon}^{(l-1)}+(2l-D+2)\partial_u\tilde{\epsilon}^{(l)}=0;\\
&[(l-1)(l-D+2)+\Delta]{\epsilon}^{(l-1)}+(2l-D+2)\partial_u\epsilon^{(l)}-2\partial_u\tilde{\epsilon}^{(l)}+(D-2l-1)\tilde{\epsilon}^{(l-1)}=0;
\label{lgc1}
\end{aligned} 
\end{equation}
for future convenience we define 
\begin{equation}\label{eq:myk}
    k^{(p=1)}(l):=(l-1)(l-D+2).
\end{equation}
For the vector gauge parameter the procedure is very similar. The wave equation reads
\begin{equation}
-2\nabla_u\nabla_r\epsilon_{\alpha}+\nabla_r\nabla_r\epsilon_{\alpha}+\frac{1}{r^2}\gamma^{ij}\nabla_i\nabla_j\epsilon_{\alpha}=0\,.
\end{equation}
Making explicit the covariant derivatives we get
\begin{equation}
\begin{aligned}
&\bigg[\partial_r^2-2\partial_u \partial_r-\frac{(D-2)}{r}(\partial_u-\partial_r)+\frac{\Delta}{r^2}\bigg]\epsilon_u=0,\\
&\bigg[\partial_r^2-2\partial_u \partial_r-\frac{(D-2)}{r}(\partial_u-\partial_r)+\frac{\Delta}{r^2}\bigg]\epsilon_r+\frac{D-2}{r^2}(\epsilon_u-\epsilon_r)-\frac{2}{r^3}\nabla^i \epsilon_i=0,\\
&\bigg[\partial_r^2-2\partial_u \partial_r-\frac{(D-4)}{r}(\partial_u-\partial_r)+\frac{\Delta}{r^2}\bigg]\epsilon_i-\frac{D-3}{r^3}\epsilon_i-\frac{2}{r}\nabla_i(\epsilon_u-\epsilon_r)=0.
\end{aligned}
\end{equation}
First we note that the equation for $\epsilon_u$ is the same as the one for the scalar gauge parameter due to the lack of connection symbols with $u$ as down index; then inserting the polyhomogeneous expansion for the other components of the vector gauge parameter we get
\begin{subequations}
	\label{A.7}
\begin{alignat}{6}
&[(l-1)(l-D+2)+\Delta]\tilde{\epsilon}_u^{(l-1)}+(2l-D+2)\partial_u\tilde{\epsilon}_u^{(l)}=0;
\label{A.7a}\\
&[(l-1)(l-D+2)+\Delta]{\epsilon}_u^{(l-1)}+(2l-D+2)\partial_u\epsilon_u^{(l)}-2\partial_u\tilde{\epsilon}_u^{(l)}\nonumber \\  
& +(D-2l-1)\tilde{\epsilon}_u^{(l-1)}=0;
\label{A.7b}\\
&[l(l+1-D)+\Delta]\tilde{\epsilon}_r^{(l-1)}+(2l-D+2)\partial_u\tilde{\epsilon}_r^{(l)}+(D-2)\tilde{\epsilon}_u^{(l-1)}-2\nabla^i\tilde{\epsilon}_i^{(l-2)}=0;
\label{A.7c}\\
&[l(l+1-D)+\Delta]{\epsilon}_r^{(l-1)}+(2l-D+2)\partial_u{\epsilon}_r^{(l)}+(D-2){\epsilon}_u^{(l-1)}-2\nabla^i{\epsilon}_i^{(l-2)}\nonumber \\ 
& +(D-2l-1)\tilde{\epsilon}_r^{(l-1)}-2\partial_u\tilde{\epsilon}_r^{(l)}=0;
\label{A.7d}\\
&[l(l-D+3)-1+\Delta]\tilde{\epsilon}_i^{(l-1)}+(2l-D+4)\partial_u\tilde{\epsilon}_i^{(l)}+2\nabla_i(\tilde{\epsilon}_r^{(l)}-\tilde{\epsilon}_u^{(l)})=0;
\label{A.7e}\\
&[l(l-D+3)-1+\Delta]\epsilon_i^{(l-1)}+(2l-D+4)\partial_u{\epsilon}_i^{(l)}+2\nabla_i({\epsilon}_r^{(l)}-{\epsilon}_u^{(l)}) \nonumber  \\ 
& +(D-2l-3)\partial_u\tilde{\epsilon}_i^{(l-1)}-2\partial_u\tilde{\epsilon}_i^{(l)}=0
\label{A.7f};
\end{alignat}
\end{subequations}
For future convenience we define 
\begin{equation}\label{eq:myks}
\begin{aligned}
    &k^{(p=2)}_u(l):=(l-1)(l-D+2);\\
    &k^{(p=2)}_r(l):=l(l+1-D);\\
    &k^{(p=2)}_i(l):=l(l-D+3).\\
\end{aligned}    
\end{equation}
Together with these equations, we have to taken into account the divergence free of the gauge parameter:
\begin{equation}
    \nabla^{\mu}\epsilon_{\mu}=-\partial_u\epsilon_r+\frac{1}{r}(r\partial_r+D-2)(\epsilon_r-\epsilon_u)+\frac{1}{r^2}\nabla^i\epsilon_i=0;
\end{equation}
which inserting the polyhomogeneous expansion gives us
\begin{equation}
\begin{aligned}
&\partial_u\tilde{\epsilon}_r^{(l)}-(l-D+1)(\tilde{\epsilon}_u^{(l-1)}-\tilde{\epsilon}_r^{(l-1)})-\nabla^i\tilde{\epsilon}_i^{(l-2)}=0;\\
&\partial_u\epsilon_r^{(l)}-(l-D+1)(\epsilon_u^{(l-1)}-{\epsilon}_r^{(l-1)})-\nabla^i\epsilon_i^{(l-2)}+\tilde{\epsilon}_u^{(l-1)}-\tilde{\epsilon}_r^{(l-1)}=0.
\label{divfree}
\end{aligned}
\end{equation}
We will see in the main text that this is enough to get answers also for the generic $p$ case. We cross-checked the above equations against the corresponding ones in Ref.~\cite{Campoleoni:2019ptc} and find agreement.

\section{The necessity of logarithms for $p= 1, 2$ } \label{Appendix-logs}
For the 1-form, let us consider the Lorenz gauge fixing preserving conditions \eqref{lgc1} and the ansatz \eqref{expansion1formg}. The recursion relations are trivially satisfied for $l\le\frac{D-6}{2}$, and for $l=\frac{D-4}{2}$, we have
\begin{equation}
    \partial_u \epsilon^{(\frac{D-4}{2})}
    =
    0,
\end{equation} 
which is obeyed since $\epsilon^{(\frac{D-4}{2})}$ only depends on the angles.
Considering $l=\frac{D-2}{2}$, we get (the symbol $k^{(p=1)}$ is defined in \eqref{eq:myk})
\begin{equation}\label{eq:epswithlog}
    \bigg[-k^{(p=1)}\bigg(\frac{D-2}{2}\bigg)+\Delta\bigg]\epsilon^{(\frac{D-4}{2})}=2\partial_u\tilde{\epsilon}^{(\frac{D-2}{2})}\,.
\end{equation}
If we suppressed the logarithmic terms, we would get
\begin{equation}
    \bigg[-k^{(p=1)}\bigg(\frac{D-2}{2}\bigg)+\Delta\bigg]\epsilon^{(\frac{D-4}{2})}=0 \Rightarrow \epsilon^{(\frac{D-4}{2})}=0\,.
\end{equation}
This mechanism, already discussed in \cite{Campoleoni:2019ptc} for the case of a one-form, would thus trivialize the charges in \eqref{charge1rad}. Instead, thanks to the presence of the logarithmic series, \eqref{eq:epswithlog} can be solved by 
\begin{equation}\label{eq:explicitsoleps}
\tilde{\epsilon}^{(\frac{D-2}{2})}(u,x)
=
\frac{u}{2}\,\bigg[-k^{(p=1)}\bigg(\frac{D-2}{2}\bigg)+\Delta\bigg]\epsilon^{(\frac{D-4}{2})}(x)
+
\hat{q}^{(\frac{D-2}{2})}(x)\,,
\end{equation}
with $\hat{q}^{(\frac{D-2}{2})}(x)$ a $u-$independent integration function. Note that the first of \eqref{lgc1} indeed does not impose any additional constraint on $\partial_u \tilde{\epsilon}^{(\frac{D-2}{2})}$. The function $\epsilon^{(\frac{D-2}{2})}(u,x)$ is instead completely free, and the remaining equations for  $l\ge\frac{D}{2}$ can be solved recursively and expressed in terms of $\epsilon^{(\frac{D-4}{2})}(x)$, $\epsilon^{(\frac{D-2}{2})}(u,x)$ and of suitable integration functions as discussed in Ref.~\cite{Campoleoni:2019ptc}.
This analysis establishes that the parameter \eqref{expansion1formg} indeed admits an arbitrary angular dependence in $\epsilon^{(\frac{D-4}{2})}(x)$ (in addition to the free function$\epsilon^{(\frac{D-2}{2})}(u,x)$).

In the same manner, for the 2-form, let us consider the Lorenz gauge fixing preserving conditions \eqref{A.7} and the divergence free condition. Before proceeding on, let us consider how to fix the gauge for gauge fixing \eqref{gxg}. The conditions to fix the gauge for gauge fixing are given by
\begin{equation}
\begin{aligned}
    &\tilde{\epsilon}^{(\frac{D}{2})}_u=-\partial_u\tilde{\epsilon}^{(\frac{D}{2})};\\
    &\epsilon^{(\frac{D-8}{2})}_i=-\partial_i\epsilon^{(\frac{D-8}{2})};\\
    &\tilde{\epsilon}^{(\frac{D-6}{2})}_i=-\partial_i\tilde{\epsilon}^{(\frac{D-6}{2})}.
    \label{gaugefixgxg}
\end{aligned}    
\end{equation}
In order to impose these constraints we need to consider a 2-polyhomogeneous expansion for the gauge for gauge parameter
\begin{equation}
    \epsilon=\sum_{l\in \frac{1}{2}\mathbb{Z}}\frac{\epsilon^{(l)}(u,x^i)+\tilde{\epsilon}^{(l)}(u,x^i)\ln(r)+\tilde{\tilde{\epsilon}}^{(l)}(u,x^i)\ln^2(r)}{r^l}.
    \label{2poly}
\end{equation}
which inserted in $\Box \epsilon=0$ gives us
\begin{subequations}
\begin{alignat}{3}
&[(l-1)(l-D+2)+\Delta]\tilde{\tilde{\epsilon}}^{(l-1)}+(2l-D+2)\partial_u\tilde{\tilde{\epsilon}}^{(l)}=0;
\label{B5a}\\
&[(l-1)(l-D+2)+\Delta]{\tilde{\epsilon}}^{(l-1)}+(2l-D+2)\partial_u{\tilde{\epsilon}}^{(l)}-4\partial_u\tilde{\tilde{\epsilon}}^{(l)} \nonumber \\ 
& +2(D-2l-1)\tilde{\tilde{\epsilon}}^{(l-1)}=0;
\label{B5b}\\
&[(l-1)(l-D+2)+\Delta]{{\epsilon}}^{(l-1)}+(2l-D+2)\partial_u{{\epsilon}}^{(l)}+2\tilde{\tilde{\epsilon}}^{(l-1)}-2\partial_u{\tilde{\epsilon}}^{(l)} \nonumber  \\ 
& +(D-2l-1)\tilde{\epsilon}^{(l-1)}=0. \label{B5c}
\end{alignat}
\label{lgc2poly}
\end{subequations}
From equation \eqref{B5a}  with $l=\frac{D-2}{2}$ and proceeding recursively we find that 
\begin{equation}
    \tilde{\tilde{\epsilon}}^{(l)}=0 \ \ \ \ \ \ \ \forall l \leq \frac{D-4}{2};
\end{equation}
moreover from equation \eqref{B5b} with $l=\frac{D+2}{2}$, $l=\frac{D}{2}$, $l=\frac{D-2}{2}$ and $l=\frac{D-4}{2}$ we get respectively
\begin{subequations}
\begin{alignat}{4}
&\bigg[-k^{(p=1)}\bigg(\frac{D+2}{2}\bigg)+\Delta\bigg]\tilde{\epsilon}^{(\frac{D}{2})}=-4\partial_u\tilde{\epsilon}^{(\frac{D+2}{2})}+4\partial_u\tilde{\tilde{\epsilon}}^{(\frac{D+2}{2})}+6\tilde{\tilde{\epsilon}}^{(\frac{D}{2})};
\label{B.7a}\\
&\bigg[-k^{(p=1)}\bigg(\frac{D}{2}\bigg)+\Delta\bigg]\tilde{\epsilon}^{(\frac{D-2}{2})}=-2\partial_u\tilde{\epsilon}^{(\frac{D}{2})}+4\partial_u\tilde{\tilde{\epsilon}}^{(\frac{D}{2})}+2\tilde{\tilde{\epsilon}}^{(\frac{D-2}{2})};
\label{B.7b}\\
&\bigg[-k^{(p=1)}\bigg(\frac{D-2}{2}\bigg)+\Delta\bigg]\tilde{\epsilon}^{(\frac{D-4}{2})}=4\partial_u\tilde{\tilde{\epsilon}}^{(\frac{D-2}{2})};
\label{B.7c}\\
&\bigg[-k^{(p=1)}\bigg(\frac{D-4}{2}\bigg)+\Delta\bigg]\tilde{\epsilon}^{(\frac{D-6}{2})}=2\partial_u\tilde{\epsilon}^{(\frac{D-4}{2})};
\label{B.7d}
\end{alignat}
\end{subequations}
while from \eqref{B5c} with $l=\frac{D+2}{2}$, $l=\frac{D}{2}$, $l=\frac{D-2}{2}$, $l=\frac{D-4}{2}$, $l=\frac{D-6}{2}$ and $l=\frac{D-8}{2}$ we get respectively
\begin{subequations}
\begin{alignat}{6}
&\bigg[-k^{(p=1)}\bigg(\frac{D+2}{2}\bigg)+\Delta\bigg]\epsilon^{(\frac{D}{2})}=-4\partial_u\epsilon^{(\frac{D+2}{2})}-2\tilde{\tilde{\epsilon}}^{(\frac{D}{2})}+2\partial_u{\tilde{\epsilon}}^{(\frac{D+2}{2})}+3{\tilde{\epsilon}}^{(\frac{D}{2})};
\label{B8a}\\
&\bigg[-k^{(p=1)}\bigg(\frac{D}{2}\bigg)+\Delta\bigg]\epsilon^{(\frac{D-2}{2})}=-2\partial_u\epsilon^{(\frac{D}{2})}-2\tilde{\tilde{\epsilon}}^{(\frac{D-2}{2})}+2\partial_u{\tilde{\epsilon}}^{(\frac{D}{2})}+{\tilde{\epsilon}}^{(\frac{D-2}{2})};
\label{B8b}\\
&\bigg[-k^{(p=1)}\bigg(\frac{D-2}{2}\bigg)+\Delta\bigg]\epsilon^{(\frac{D-4}{2})}=2\partial_u{\tilde{\epsilon}}^{(\frac{D-2}{2})}-{\tilde{\epsilon}}^{(\frac{D-4}{2})};
\label{B8c}\\
&\bigg[-k^{(p=1)}\bigg(\frac{D-4}{2}\bigg)+\Delta\bigg]\epsilon^{(\frac{D-6}{2})}=2\partial_u\epsilon^{(\frac{D-4}{2})}+2\partial_u{\tilde{\epsilon}}^{(\frac{D-4}{2})}-3{\tilde{\epsilon}}^{(\frac{D-6}{2})};
\label{B8d}\\
&\bigg[-k^{(p=1)}\bigg(\frac{D-6}{2}\bigg)+\Delta\bigg]\epsilon^{(\frac{D-8}{2})}=4\partial_u\epsilon^{(\frac{D-6}{2})}+2\partial_u{\tilde{\epsilon}}^{(\frac{D-6}{2})}-5{\tilde{\epsilon}}^{(\frac{D-8}{2})};
\label{B8e}\\
&\bigg[-k^{(p=1)}\bigg(\frac{D-8}{2}\bigg)+\Delta\bigg]\epsilon^{(\frac{D-10}{2})}=6\partial_u\epsilon^{(\frac{D-8}{2})}+2\partial_u{\tilde{\epsilon}}^{(\frac{D-8}{2})}-7{\tilde{\epsilon}}^{(\frac{D-10}{2})}.
\label{B8f}
\end{alignat}
\end{subequations}
The gauge for gauge fixing equation fix for us the functions $\tilde{\epsilon}^{(\frac{D}{2})}$, $\tilde{\epsilon}^{(\frac{D-6}{2})}$ and $\epsilon^{(\frac{D-8}{2})}$;  
then equations \eqref{B.7c} and \eqref{B.7d} fix the functions $\tilde{\tilde{\epsilon}}^{(\frac{D-2}{2})}$ and $\tilde{\epsilon}^{(\frac{D-4}{2})}$. Moreover, considering equation \eqref{B5b} for $l\leq \frac{D-6}{2}$ we get that all the functions $\tilde{\epsilon}^{(l)}$ with $l \leq \frac{D-8}{2}$ are fixed. At this point, equations \eqref{B8c}, \eqref{B8d}, \eqref{B8e} and  \eqref{B8f} together to \eqref{B5c} for $l \leq \frac{D-10}{2}$ fix the functions $\epsilon^{(\frac{D-6}{2})}$, $\epsilon^{(\frac{D-4}{2})}$, $\tilde{\epsilon}^{(\frac{D-2}{2})}$, and $\epsilon^{(l)}$ with $l\leq \frac{D-10}{2}$. However, from equations \eqref{B.7a}, \eqref{B.7b}, \eqref{B8a} and \eqref{B5b} we have the unfixed functions $\tilde{\tilde{\epsilon}}^{(\frac{D+2}{2})}$, $\tilde{\tilde{\epsilon}}^{(\frac{D}{2})}$, ${\tilde{\epsilon}}^{(\frac{D+2}{2})}$, ${\tilde{\epsilon}}^{(\frac{D-2}{2})}$, $\epsilon^{(\frac{D+2}{2})}$, $\epsilon^{(\frac{D}{2})}$ and $\epsilon^{(\frac{D-2}{2})}$ to stop the chain of fixing functions and be sure not to set gauge fixing with the same arbitrary functions.

After these considerations on the gauge fixing, we can proceed considering the preserving Lorenz gauge condition for the 2-form. Starting from \eqref{A.7a}, we consider first $l=\frac{D}{2}$ getting
\begin{equation}
    \bigg[-k_u^{(p=2)}\bigg(\frac{D}{2}\bigg)+\Delta \bigg]\tilde{\epsilon}_u^{(\frac{D-2}{2})}=0 \Rightarrow \tilde{\epsilon}_u^{(\frac{D-2}{2})}=0;
\end{equation}
next, we consider $l=\frac{D-2}{2}$ to get
\begin{equation}
    \bigg[-k_u^{(p=2)}\bigg(\frac{D-2}{2}\bigg)+\Delta \bigg]\tilde{\epsilon}_u^{(\frac{D-4}{2})}=0 \Rightarrow \tilde{\epsilon}_u^{(\frac{D-4}{2})}=0;
\end{equation}
going on in this way we get
\begin{equation}
    \tilde{\epsilon}_u^{(l)}=0 \ \ \ \ \ \ \ \forall l \leq \frac{D}{2},
\end{equation}
where the equality rises from our gauge for gauge fixing. Now we take into account equation \eqref{A.7b} and we consider respectively $l=\frac{D-2}{2}$ and $l=\frac{D-4}{2}$:
\begin{equation}
\begin{aligned}
    &\bigg[-k_u^{(p=2)}\bigg(\frac{D-2}{2}\bigg)+\Delta \bigg]{\epsilon}_u^{(\frac{D-4}{2})}=0 \Rightarrow {\epsilon}_u^{(\frac{D-4}{2})}=0;\\
    &\bigg[-k_u^{(p=2)}\bigg(\frac{D-4}{2}\bigg)+\Delta \bigg]{\epsilon}_u^{(\frac{D-6}{2})}=0 \Rightarrow {\epsilon}_u^{(\frac{D-6}{2})}=0;\\
\end{aligned}
\end{equation}
going on, we get
\begin{equation}
    {\epsilon}_u^{(l)}=0 \ \ \ \ \ \ \ \forall l \leq \frac{D-4}{2},
\end{equation}
It is now the turn of \eqref{A.7c}; let us start with $l=\frac{D-2}{2}$, we get
\begin{equation}
    \bigg[-k_r^{(p=2)}\bigg(\frac{D-2}{2}\bigg)+\Delta \bigg]\tilde{\epsilon}_r^{(\frac{D-4}{2})}=0 \Rightarrow \tilde{\epsilon}_r^{(\frac{D-4}{2})}=0;
\end{equation}
if we now consider equation \eqref{A.7c} with $l=\frac{D-4}{2}$ and equation \eqref{A.7e} with $l=\frac{D-6}{2}$ we get 
\begin{equation}
\begin{aligned}
    &\bigg[-k^{(p=2)}_r\bigg(\frac{D-4}{2}\bigg)+\Delta \bigg]\tilde{\epsilon}_r^{(\frac{D-6}{2})}=2\nabla^i\tilde{\epsilon}_i^{(\frac{D-8}{2})};\\
    &\bigg[-k^{(p=2)}_i\bigg(\frac{D-6}{2}\bigg)+\Delta \bigg]\tilde{\epsilon}_i^{(\frac{D-8}{2})}=-2\nabla^i\tilde{\epsilon}_r^{(\frac{D-6}{2})};
    \label{sis1}
\end{aligned}    
\end{equation}
and using results from Appendix~\ref{app3} we find that 
\begin{equation}
    \tilde{\epsilon}_r^{(\frac{D-6}{2})}=\tilde{\epsilon}_i^{(\frac{D-8}{2})}=0.
\end{equation}
Taking now equation \eqref{A.7c} with $l=\frac{D-6}{2}$ and equation \eqref{A.7e} with $l=\frac{D-8}{2}$ we get 
\begin{equation}
\begin{aligned}
    &\bigg[-k^{(p=2)}_r\bigg(\frac{D-6}{2}\bigg)+\Delta \bigg]\tilde{\epsilon}_r^{(\frac{D-8}{2})}=2\nabla^i\tilde{\epsilon}_i^{(\frac{D-10}{2})};\\
    &\bigg[-k^{(p=2)}_i\bigg(\frac{D-8}{2}\bigg)+\Delta \bigg]\tilde{\epsilon}_i^{(\frac{D-10}{2})}=-2\nabla^i\tilde{\epsilon}_r^{(\frac{D-8}{2})};
    \label{sis2}
\end{aligned}    
\end{equation}
and using again results from Appendix~\ref{app3} we find that 
\begin{equation}
    \tilde{\epsilon}_r^{(\frac{D-8}{2})}=\tilde{\epsilon}_i^{(\frac{D-10}{2})}=0.
\end{equation}
Going on we get
\begin{equation}
\begin{aligned}
    &\tilde{\epsilon}_r^{(l)}=0 \ \ \ \forall \ l \leq \frac{D-4}{2};\\
    &\tilde{\epsilon}_i^{(l)}=0 \ \ \ \forall \ l \leq \frac{D-6}{2}
    \end{aligned}
\end{equation}
where the equality in the second equation take into account our gauge for gauge fixing.
From the conditions that falloffs are preserved up  to logarithmic terms one gets
\begin{equation}
\begin{aligned}
    &\partial_u\epsilon_r^{(l)}=0 \ \ \ \forall \ l \leq \frac{D-4}{2},\\
    &\partial_u\epsilon_i^{(l)}=0 \ \ \ \forall \ l \leq \frac{D-6}{2}.\\
\end{aligned}
\end{equation}
Now, considering the first equation \eqref{divfree} with $l=\frac{D-2}{2}$ we get
\begin{equation}
    \partial_u\tilde{\epsilon}_r^{(\frac{D-2}{2})}=0.
    \label{utr}
\end{equation}
Let us consider equation \eqref{A.7d} with $l=\frac{D-4}{2}$ we have
\begin{equation}
    \bigg[-k_r^{(p=2)}\bigg(\frac{D-4}{2}\bigg)+\Delta \bigg]{\epsilon}_r^{(\frac{D-6}{2})}=0 \Rightarrow {\epsilon}_r^{(\frac{D-6}{2})}=0;
\end{equation}
considering, now equation \eqref{A.7d} with $l=\frac{D-6}{2}$ and equation \eqref{A.7f} with $l=\frac{D-8}{2}$ we get 
\begin{equation}
\begin{aligned}
    &\bigg[-k^{(p=2)}_r\bigg(\frac{D-6}{2}\bigg)+\Delta \bigg]{\epsilon}_r^{(\frac{D-8}{2})}=2\nabla^i{\epsilon}_i^{(\frac{D-10}{2})};\\
    &\bigg[-k^{(p=2)}_i\bigg(\frac{D-8}{2}\bigg)+\Delta \bigg]{\epsilon}_i^{(\frac{D-10}{2})}=-2\nabla^i{\epsilon}_r^{(\frac{D-8}{2})};
    \label{sis3}
\end{aligned}    
\end{equation}
and using results from Appendix~\ref{app3} we find that 
\begin{equation}
    {\epsilon}_r^{(\frac{D-8}{2})}={\epsilon}_i^{(\frac{D-10}{2})}=0;
\end{equation}
Like the case of tilded functions above, we can go on using results of Appendix~\ref{app3}; therefore
\begin{equation}
\begin{aligned}
    &{\epsilon}_r^{(l)}=0 \ \ \ \forall \ l \leq \frac{D-6}{2};\\
    &{\epsilon}_i^{(l)}=0 \ \ \ \forall \ l \leq \frac{D-8}{2}.
    \end{aligned}
\end{equation}
The  fifth and the seventh equation of the preserving fall-off condition up to logarithm are now trivially solved.\\
Let us now show the necessity of logarithmic terms for the 2-form; considering equation \eqref{A.7d} with $l=\frac{D-2}{2}$ and equation \eqref{A.7f} with $l=\frac{D-4}{2}$ we get, using constrain \eqref{utr}
\begin{equation}
\begin{aligned}
    &\bigg[-k^{(p=2)}_r\bigg(\frac{D-2}{2}\bigg)+\Delta \bigg]{\epsilon}_r^{(\frac{D-4}{2})}=2\nabla^i{\epsilon}_i^{(\frac{D-6}{2})};\\
    &\bigg[-k^{(p=2)}_i\bigg(\frac{D-4}{2}\bigg)+\Delta \bigg]{\epsilon}_i^{(\frac{D-6}{2})}=-2\nabla^i{\epsilon}_r^{(\frac{D-4}{2})}+2\partial_u\tilde{\epsilon}_i^{(\frac{D-4}{2})};
    \label{sis4}
\end{aligned}    
\end{equation}
and we would deduce that ${\epsilon}_r^{(\frac{D-4}{2})}={\epsilon}_i^{(\frac{D-6}{2})}=0$ due to results in Appendix~\ref{app3} if logarithmic terms in the gauge parameter expansion would suppressed: the presence of $\tilde{\epsilon}_i^{(\frac{D-4}{2})}$ is the lifesaver of asymptotic symmetry since it indirectly makes the asymptotic charges non-zero.\\
The very same analysis, with minimal modification, can be retraced in the case of Coulomb fall-offs leading to the same results.

\section{Asymptotic solution space and matter couplings}\label{Appendix:Physical_logs}
In our analysis of residual symmetries in Lorenz gauge we are forced to include logarithmic falloffs on the gauge parameters. In the bulk of our presentation we treated them as pure-large gauge transformations, that as such do not affect the  large-$r$ behaviour of field-strength components.
In this section we consider the possibility that they do modify the falloffs of physical fields and analyse their consequences in the presence of matter couplings, in the spirit of 
\cite{Himwich:2019qmj, Peraza:2023ivy}.

Let us first consider the case of scalar QED. We assume that the scalar matter displays the leading radial behaviour 
\begin{equation}
    \phi \sim \frac{\phi^{(\frac{D-2}{2})}}{r^{\frac{D-2}{2}}},
    \label{scalarasy}
\end{equation} 
so that the falloffs of the current
$J_{\mu}=ie\phi D_{\mu}\phi$ are given by
 \begin{equation}
    J_u \sim \frac{J_u^{(D-2)}}{r^{D-2}}, \ \ \ J_r \sim \frac{J_r^{(D-1)}}{r^{D-1}}, \ \ \ J_i \sim \frac{J_i^{(D-2)}}{r^{D-2}} .
\end{equation}
Thus, the equations of motion in Lorenz gauge,
\begin{equation}
    \Box B_{\mu}=J_{\mu},
\end{equation}
can be solved as in Appendix~\ref{Appendix:eqs} with the difference  that the subleading orders satisfy an equation with source. We consider the following polyhomogeneous expansions for the gauge field
\begin{equation}
\begin{aligned}
    B_u &\sim \frac{B_u^{(\frac{D-2}{2})}}{r^{\frac{D-2}{2}}}+\frac{\tilde{B}_u^{(\frac{D-2}{2})}\ln(r)}{r^{\frac{D-2}{2}}}+\cdots ,\\ B_r &\sim \frac{B_r^{(\frac{D-2}{2})}}{r^{\frac{D-2}{2}}}+\frac{B_r^{(\frac{D}{2})}}{r^{\frac{D}{2}}}+\frac{\tilde{B}_r^{(\frac{D}{2})}\ln(r)}{r^{\frac{D}{2}}}+\cdots , \\ B_i &\sim \frac{B_i^{(\frac{D-4}{2})}}{r^{\frac{D-4}{2}}}+\frac{\tilde{B}_i^{(\frac{D-2}{2})}\ln(r)}{r^{\frac{D-2}{2}}}+\cdots  .\ \
\end{aligned}    
\end{equation}
Looking at the equation for $\mu=r$ for the logarithmic branch, i.e. equation \eqref{A.7c}, we get for $l=\frac{D}{2}$
\begin{equation} \label{log1}
    2\partial_u \tilde{B}_r^{(\frac{D}{2})}+(D-2)\tilde{B}_u^{(\frac{D-2}{2})}=0;
\end{equation}
while for the Lorenz gauge condition of the non-logarithmic branch, i.e. the second of equations \eqref{divfree}, we get with $l=\frac{D-2}{2}$
\begin{equation} \label{log2}
    \partial_u B_r^{(\frac{D-2}{2})}=0.
\end{equation}
The relations \eqref{log1} and \eqref{log2} are instrumental to show in particular that the logs do not modify the leading behaviour of the component $H_{ru}$ of the field strength, since
\begin{equation}
\begin{split}
    H_{ru} \, \sim \,  &\frac{\tilde{B}_u^{(\frac{D-2}{2})}}{r^{\frac{D}{2}}}-\bigg(\frac{D-2}{2}\bigg)\frac{B_u^{(\frac{D-2}{2})}}{r^{\frac{D}{2}}}-\bigg(\frac{D-2}{2}\bigg)\frac{\tilde{B}_u^{(\frac{D-2}{2})}\ln(r)}{r^{\frac{D}{2}}} \\
    & -\frac{\partial_u B_r^{(\frac{D-2}{2})}}{r^{\frac{D-2}{2}}}-\frac{\partial_u B_r^{(\frac{D}{2})}}{r^{\frac{D}{2}}}-\frac{\partial_u \tilde{B}_r^{(\frac{D}{2})}\ln(r)}{r^{\frac{D}{2}}} \sim \mathcal{O}(r^{-\frac{D}{2}}),
\end{split}
\end{equation}
thus ensuring the finiteness of energy flux at null infinity. 
%\CH{Sono confuso, il calcolo della carica coinvolge $H_{ur}$, ma quello del flusso di energia coinvolge un oggetto quadratico nella field strength (che in particolare dev'essere sensibile alle componenti angolari).}

We can repeat the same analysis for $p-$forms.  Let us consider the wave equation with source
\begin{equation}
    \Box B_{\mu_1\cdots \mu_{p}}=J_{\mu_1\cdots \mu_{p}}
\end{equation}
where the leading falloffs of the matter part are assumed to be
\begin{equation}
     J_{ri_1\cdots i_{p-1}} \sim \frac{J_r^{(\frac{D-2p+2}{2})}}{r^{\frac{D-2p+2}{2}}}.
\end{equation}
Currents with this type of behaviour can arise, for instance, through a coupling of a $p-$form to a scalar field with the vertex $\mathcal{V}_{\mathcal{N}=0}=e^{k\phi}H_{\mu_1\cdots \mu_{p+1}}H^{\mu_1\cdots \mu_{p+1}}$.
We consider a polyhomogeneous expansion  with radiation leading fall-offs
\begin{equation} \label{ansatzp}    
\begin{aligned}
    B_{ui_1\cdots i_{p-1}} &\sim \frac{B_{ui_1\cdots i_{p-1}}^{(\frac{D-2p}{2})}}{r^{\frac{D-2p}{2}}}+\frac{\tilde{B}_{ui_1\cdots i_{p-1}}^{(\frac{D-2p}{2})}\ln(r)}{r^{\frac{D-2p}{2}}}+\cdots ,\\ B_{rui_1\cdots i_{p-2}} &\sim \frac{B_{rui_1\cdots i_{p-2}}^{(\frac{D-2p+2}{2})}}{r^{\frac{D-2p+2}{2}}}+\frac{B_{rui_1\cdots i_{p-2}}^{(\frac{D-2p+4}{2})}}{r^{\frac{D-2p+4}{2}}}+\frac{\tilde{B}_{rui_i\cdots i_{p-2}}^{(\frac{D-2p+4}{2})}\ln(r)}{r^{\frac{D-2p+4}{2}}}+\cdots , \\ B_{ri_1\cdots i_{p-1}} &\sim \frac{B_{ri_1\cdots i_{p-1}}^{(\frac{D-2p}{2})}}{r^{\frac{D-2p}{2}}}+\frac{B_{ri_1\cdots i_{p-1}}^{(\frac{D-2p+2}{2})}}{r^{\frac{D-2p+2}{2}}}+\frac{\tilde{B}_{ri_1\cdots i_{p-1}}^{(\frac{D-2p+2}{2})}\ln(r)}{r^{\frac{D-2p+2}{2}}}+\cdots , \\ B_{i_i\cdots i_p} &\sim \frac{B_{i_i..i_p}^{(\frac{D-2p-2}{2})}}{r^{\frac{D-2p-2}{2}}}+\frac{\tilde{B}_{i_1\cdots i_p}^{(\frac{D-2p}{2})}\ln(r)}{r^{\frac{D-2p}{2}}}+\cdots , \ \
\end{aligned}    
\end{equation}
where we took into account both the Lorenz gauge and the equations of motion. 

 Let us start with the Lorenz-like gauge condition $\nabla^{\nu}B_{\nu \mu_1...\mu_{p-1}}=0$ for $(\mu_1,...,\mu_{p-1})=(i_1,...,i_{p-1})$, inserting a generic polyhomogeneous expansion for the gauge field components we get
\begin{equation} \label{divfreepform}
\begin{aligned}
&\partial_u\tilde{B}_{ri_1...i_{p-1}}^{(l)}-G_{p,D}(\tilde{B}_{u{i_1...i_{p-1}}}^{(l-1)}-\tilde{B}_{ri_1...i_{p-1}}^{(l-1)})-\nabla^i\tilde{B}_{ii_1...i_{p-1}}^{(l-2)}=0;\\
&\partial_u{B}_{ri_1...i_{p-1}}^{(l)}-G_{p,D}({B}_{u{i_1...i_{p-1}}}^{(l-1)}-{B}_{ri_1...i_{p-1}}^{(l-1)})-\nabla^i{B}_{ii_1...i_{p-1}}^{(l-2)}+\tilde{B}_{ui_1...i_{p-1}}^{(l-1)}-\tilde{B}_{ri_1...i_{p-1}}^{(l-1)}=0.
\end{aligned}
\end{equation}
where $G_{p,D}:=(l-D+2p-1)$. Choosing $l=\frac{D-2p}{2}$ from the Lorenz gauge condition for purely angular components and considering the expansions \eqref{ansatzp} we obtain
\begin{equation}\label{lorenzgaugeconndrad}
    \partial_u{B}_{ri_1\cdots i_{p-1}}^{(\frac{D-2p}{2})}=0.
\end{equation}
The homogeneous equation of motion with $(\mu_1,\ldots,\mu_p)=(r,i_1,\ldots,i_{p-1})$ looks
\begin{equation}
\begin{aligned}
\mathcal{D}_{p,D}B_{ri_1\cdots i_{p-1}}+\frac{E_{p,D}^{(1)}}{r^2}B_{ri_1\cdots i_{p-1}}+\frac{E_{p,D}^{(2)}}{r^2}B_{ui_1\cdots i_{p-1}}+\frac{\sum B}{r}-\frac{2}{r^3}\nabla^iB_{ii_1\cdots i_{p-1}}=0.
\end{aligned}
\end{equation}
where 
\begin{equation}
\begin{aligned}
&\mathcal{D}_{p,D}:=\partial_r^2-2\partial_u\partial_r-\frac{(D-2p)}{r}(\partial_u-\partial_r)+\frac{\Delta}{r^2}; \\ 
&E^{(1)}_{p,D}:=p^2-2p+3-D; \\ 
&E^{(2)}_{p,D}:=D-2p;\\
&\sum B:=\sum_{s=1}^{p-1}(-1)^{s-1}\partial_{i_s}B_{rui_1\cdots i_{s-1}i_{s+1}\cdots i_{p-1}}
\end{aligned}
\end{equation}
Inserting a generic polyhomogeneous expansion, we get the order by order relations
\begin{equation} \label{orbyorpr}
\begin{aligned}
&[k_r^{(p)}(l)+\Delta]\tilde{B}^{(l-1)}_{ri_1\cdots i_{p-1}}+A_1\partial_u\tilde{B}^{(l)}_{ri_1\cdots i_{p-1}}+A_2 \tilde{B}^{(l-1)}_{ui_1\cdots i_{p-1}}+\sum \tilde{B}^{(l-1)}-2\nabla^i\tilde{B}^{(l-2)}_{ii_1\cdots i_{p-1}}=0;\\
&[k_r^{(p)}(l)+\Delta]B^{(l-1)}_{ri_1\cdots i_{p-1}}+A_1\partial_uB^{(l)}_{ri_1\cdots i_{p-1}}+A_2 B^{(l-1)}_{ui_1\cdots i_{p-1}}+\sum B^{(l-1)}-2\nabla^iB^{(l-2)}_{ii_1\cdots i_{p-1}} \\
&+A_3\tilde{B}^{(l-1)}_{ri_1\cdots i_{p-1}}-2\partial_u\tilde{B}^{(l)}_{ri_1\cdots i_{p-1}}=0;
\end{aligned}
\end{equation}
where 
\begin{equation}
\begin{aligned}
&k_r^{(p)}(l):=l(l-1-D+2p)-4p+p^2+3;\\
&A_1:=2l-D+2p;\\
&A_2:=D-2p;\\
&A_3:=D-2l+2p-1;\\
&\sum \tilde{B}^{(l-1)}:=\sum_{s=1}^{p-1}(-1)^{s}\partial_{i_s}\tilde{B}^{(l-1)}_{rui_1\cdots i_{s-1}i_{s+1}\cdots i_{p-1}};\\
&\sum B^{(l-1)}:=\sum_{s=1}^{p-1}(-1)^{s}\partial_{i_s}B^{(l-1)}_{rui_1\cdots i_{s-1}i_{s+1}\cdots i_{p-1}}. \label{orbyorprln}
\end{aligned}   
\end{equation} 
Now, taking the current falloffs into account, the first order that satisfy the first of \eqref{orbyorprln} is $l=\frac{D-2p+2}{2}$, given that we have to cancel out the order $\mathcal{O}(r^{-\frac{D-2p+2}{2}}\ln(r))$. Therefore, inserting $l=\frac{D-2p+2}{2}$ in \eqref{orbyorprln} and using the expansions \eqref{ansatzp} we get
\begin{equation}\label{eqmotcondrad}
2\partial_u\tilde{B}_{ri_1\cdots i_{p-1}}^{(\frac{D-2p+2}{2})}+(D-2p)\tilde{B}_{ui_1\cdots i_{p-1}}^{(\frac{D-2p}{2})}=0.
\end{equation}
Thus, in same manner as for the case $p=1$, there is a cancellation of the potentially dangerous leading order logarithms in the field strength component $H_{rui_1\cdots i_{p-1}}$ that guarantees finiteness of the energy flux. In the case of a polyhomogeneous expansion for the gauge field components with Coulomb fall-offs, equations \eqref{lorenzgaugeconndrad} and \eqref{eqmotcondrad} are modified to
\begin{equation}\label{eqmotlorcondcoul}
\begin{aligned}
    &\partial_u{B}_{ri_1\cdots i_{p-1}}^{(D-2p-1)}=0;\\
    &(D-2p)\big[\partial_u\tilde{B}_{ri_1\cdots i_{p-1}}^{(D-2p)}+\Tilde{B}_{ui_1\cdots i_{p-1}}^{(D-2p-1)}\big]=0.
\end{aligned}    
\end{equation}

\section{A useful result}\label{app3}
Let us consider the following differential equation system
\begin{subequations}
\begin{alignat}{2}
    &[-k_1+\Delta]f-2\nabla^ih_i=0;
        \label{C1a}\\
    &[-k_2+\Delta]h_i+2\nabla_if=0;
    \label{C1b}
\end{alignat}
\end{subequations}
where $k_1$ and $k_2$ are parametric positive numbers while $f$ and $h_i$ sufficiently smooth arbitrary functions. This kind of system is the one appearing in \eqref{sis1}, \eqref{sis2}, \eqref{sis3} and, if logarithmic terms would suppressed, in \eqref{sis4}. \\
Taking the gradient of equation \eqref{C1b} we get
\begin{equation}
    2\Delta f-k_2\nabla^ih_i+\nabla^i\nabla^j\nabla_j h_i=0;
    \label{C2}
\end{equation}
where, due to curvature effects, we cannot exchange covariant derivatives. However using the Riemann tensor components identity $[\nabla^i,\nabla^j]h_i=\nabla^i\nabla^jh_i-\nabla^j\nabla^ih_i=R^k_{iij}h_k=g^{ks}R_{siij}h_j$ we can go on. Indeed
\begin{equation}
\begin{aligned}
\nabla^i\nabla^j(\nabla_jh_i)&=\nabla^j\nabla^i(\nabla_jh_i)+R^j_{kij}(\nabla^kh^i)+R^i_{kij}(\nabla^jh^k) \\
&=R^j_{kij}(\nabla^kh^i)+R^i_{kij}(\nabla^jh^k)+\nabla^j\nabla_j\nabla_ih^i+\nabla^j(R^i_{kij}h^k) \\
&=\nabla^j\nabla_j\nabla_ih^i-R_{ki}(\nabla^kh^i)+R_{kj}(\nabla^jh^k)+\nabla^j(R_{kj}h^k) \\
&=\nabla^j\nabla_j\nabla_ih^i+\nabla^j(R_{kj}h^k);
\end{aligned}    
\end{equation}
but the Riemann tensor components for a maximally symmetric space are given by
\begin{equation}
    R_{\alpha \beta \gamma \delta}=g_{\alpha \gamma}g_{\beta \delta}-g_{\alpha \delta}g_{\beta \gamma}
\end{equation}
from which, using the metric of the $(D-2)-$sphere $\gamma_{\alpha \beta}$ we get
\begin{equation}
    R_{\beta \delta}=\gamma^{\alpha \gamma} R_{\alpha \beta \gamma \delta}=(D-2)\gamma_{\beta \delta}-\delta^{\gamma}_{\delta}\gamma_{\beta \gamma}=(D-3)\gamma_{\beta \delta}.
\end{equation}
Therefore 
\begin{equation}
\nabla^i\nabla^j(\nabla_jh_i)=\nabla^j\nabla_j\nabla^ih_i+(D-3)\nabla^ih_i.
\end{equation}
Hence, returning to equation \eqref{C2} we can write
\begin{equation}
    2\Delta f+[\Delta+(D-3-k_2)]\nabla^ih_i=0;
\end{equation}
from which
\begin{equation}
    \Delta f=-\frac{1}{2}[\Delta+\Bar{k}_2]\nabla^ih_i,
    \label{soldf}
\end{equation}
where we defined $\Bar{k}_2:=(D-3-k_2)$.
Now, inserting the expression for $\Delta f$ from above in equation \eqref{C1a} we get
\begin{equation}
    -\frac{1}{2}[\Delta+\Bar{k}_2]\nabla^ih_i=2\nabla^ih_i+k_1f;
\label{eqf}    
\end{equation}
from which 
\begin{equation}
    f=-\frac{[\Delta+\Bar{k}_2+4]}{2k_1}\nabla^ih_i,
\label{solf}    
\end{equation}
and reinserting it back in the original equation we have
\begin{subequations}
\footnotesize
\begin{alignat}{1}
&-[-k_1+\Delta]\frac{[\Delta+\Bar{k}_2+4]}{2k_1}\nabla^ih_i-2\nabla^ih_i=0 \Rightarrow [\Delta^2+(\Bar{k}_2-k_1+4)\Delta-k_1\Bar{k}_2]\nabla^ih_i=0.
\end{alignat}
\end{subequations}
However, in similar way, we can find from equation \eqref{C1a}
\begin{equation}
    \nabla^ih_i=\frac{1}{2}[-k_1+\Delta]f;
\end{equation}
and substituting it in equation \eqref{C2} to get
\begin{equation}
    2\Delta f=-\frac{1}{2}[\Delta+\Bar{k}_2][-k_1+\Delta]f \Rightarrow [\Delta^2+(\Bar{k}_2-k_1+4)\Delta-k_1\Bar{k}_2]f=0.
\end{equation}
Therefore functions $\nabla^ih_i$ and $f$ satisfy the same differential equation; moreover the full biharmonic equation on the ball under sufficiently smooth hypothesis of the functions admits a unique solution. Therefore we conclude that 
\begin{equation}
    f=\nabla^ih_i;
\end{equation}
and using this information in equation \eqref{C1a} we have
\begin{equation}
    [-k_1-2+\Delta]f=0 \Rightarrow f=0,
\end{equation}
and using it in equation \eqref{C1b} we get
\begin{equation}
    [-k_2+\Delta]h_i=0 \Rightarrow h_i=0.
    \label{h}
\end{equation}
In the steps above if $k_1$ is null the equality \eqref{solf} is not defined and an alternative way is needed. We can start again form equation \eqref{soldf} and \eqref{C1a} with $k_1=0$ to get
\begin{equation}
[\Delta^2+(\Bar{k}_2+4)\Delta]f=0;
\end{equation}
while form the laplacian of \eqref{eqf} with $k_1=0$ we get
\begin{equation}
    [\Delta^2+(\Bar{k}_2+4)\Delta]\nabla^ih_i=0.
\end{equation}
Hence, again
\begin{equation}
    f=\nabla^ih_i,
\end{equation}
and from \eqref{C1a} with $k_1=0$ we deduce that
\begin{equation}
    [-2+\Delta]f=0 \Rightarrow f=0,
\end{equation}
and using it in \eqref{C1b} we get the same result in \eqref{h}.
If $k_2=0$ it is enough to use $\Bar{k}_2=D-3$ in all steps.

\providecommand{\href}[2]{#2}\begingroup\raggedright\endgroup

\end{document}